\documentclass[12pt]{article}
\usepackage{epsfig,amssymb,amsmath,psfrag}

\def \bl  {\begin{align*}}
\def \el  {\end{align*}}

\def \be  {\begin{equation}}
\def \ee  {\end{equation}}
\def \ba  {\begin{eqnarray}}
\def \ea  {\end{eqnarray}}
\def \baa {\begin{eqnarray*}}
\def \eaa {\end{eqnarray*}}
\def \bb  {\begin {thebibliography} }
\def \eb  {\end{thebibliography}}
\def \lab #1 {\label{#1}}

\def \matrix #1 {\left(\begin{array}{cc} #1 \end{array}\right)}

\renewcommand{\a}{\alpha}
\newcommand{\adt}{{\dot{\alpha}}}

\newcommand{\lam}{\lambda}
\newcommand{\tlam}{\tilde{\lambda}}
\renewcommand{\AA}{\mathcal{A}}
\newcommand{\BB}{\mathcal{B}}
\newcommand{\CC}{\mathcal{C}}
\newcommand{\cZ}{\mathcal{Z}}
\newcommand{\cW}{\mathcal{W}}

\newcommand \widebar [1] {\overline{#1}}

\def\XXint#1#2#3{{\setbox0=\hbox{$#1{#2#3}{\int}$}
     \vcenter{\hbox{$#2#3$}}\kern-.5\wd0}}

\def\l<{\langle}\def\r>{\rangle}

\begin{document}

\thispagestyle{empty}

\begin{flushright}
CERN-PH-TH/2011-151\\
LAPTH-020/11
\end{flushright}
\vskip2.2truecm
\begin{center}
\vskip 0.2truecm {\Large\bf
{\Large Tree-level amplitudes and dual superconformal symmetry}
}\\
\vskip 1truecm
{\bf J.~M. Drummond$^{}$  \\
}
\vskip 0.5cm

\begingroup
\textit{PH-TH Division, CERN, CH-1211, Geneva 23, Switzerland 
 }\\
\par
\vskip 0.5cm
\textit{
LAPTH, Universit\'e de Savoie, CNRS,\\
B.P. 110, F-74941 Annecy-le-Vieux Cedex, France}
\par
\texttt{drummond@lapp.in2p3.fr\phantom{\ldots}}
\endgroup


\end{center}

\vskip 1truecm 
\centerline{\bf Abstract} 
We review the structure of gauge theory scattering amplitudes at tree level and describe how a compact expression can be found which encodes all the tree-level amplitudes in the maximally supersymmetric $\mathcal{N}=4$ theory. The expressions for the amplitudes reveal a dual superconformal symmetry. We describe how these ideas can be extended to leading singularities and the loop integrand in the planar theory and discuss the appearance of dual conformal symmetry in higher-dimensional gauge theories. This article is an invited review for a special issue of Journal of Physics A devoted to ``Scattering Amplitudes in Gauge Theories".

\medskip

 \noindent

\newpage
\setcounter{page}{1}\setcounter{footnote}{0}


\section{Introduction}

Scattering amplitudes in gauge theories reveal surprising and novel features which are difficult to see in their initial Lagrangian formulation. Many expressions for amplitudes may be constructed without appealing directly to standard perturbative techniques involving a summation over Feynman diagrams. In particular methods relying on analytic behaviour of the on-shell amplitudes only have been developed which are very efficient (see e.g. \cite{Bern:1994zx,Bern:1994cg,Britto:2004nc,Bern:2011qt}). We will focus on methods for tree-level amplitudes in this article, in particular the idea of BCFW recursion \cite{Britto:2004ap,Britto:2005fq}.

As well as revealing surprising simplicity in the expressions for the amplitudes, the maximally supersymmetric theory in four dimensions also has a very large amount of symmetry. The theory is superconformally invariant and this fact expresses itself in the form of the tree-level amplitudes \cite{Witten:2003nn}. In addition there is another copy of superconformal symmetry, called `dual superconformal symmetry' which appears at the level of the colour-ordered partial amplitudes \cite{Drummond:2008vq}. The two copies of superconformal symmetry partially overlap and their closure is in fact the Yangian of the superconformal algebra \cite{Drummond:2009fd}. The appearance of a second copy of the superconformal symmetry is also visible in the AdS sigma model which is believed to describe the $\mathcal{N}=4$ theory at strong coupling via the AdS/CFT correspondence \cite{Berkovits:2008ic,Beisert:2008iq}. Its relation to the Yangian, a symmetry usually associated with quantum integrable systems, is indicative of a link to the remarkable structure governing the spectrum of anomalous  dimensions of gauge-invariant local operators (see \cite{Beisert:2010jr} for a review).

The large symmetry algebra of the theory makes itself felt not only at tree-level but also in the form of the four-dimensional planar integrand at each order in the loop expansion \cite{ArkaniHamed:2010kv}. 
As we will discuss the recursive techniques for tree-level amplitudes and for the loop integrand produce expressions which respect the full symmetry term by term.

After introducing the on-shell superspace describing the theory we will describe how the BCFW recursion relations can be systematically solved for the complete tree-level S-matrix of the $\mathcal{N}=4$ theory. The expressions we obtain will lead us naturally to the notion of dual superconformal symmetry. We will then discuss how this symmetry, combined with the original Lagrangian superconformal symmetry appears at the level of the loop integrand, in particular forcing all leading singularities to be Yangian invariants. These quantities can be expressed in terms of residues of a particular integral over a Grassmannian space. These ideas then combine to the idea of a recursive relation for the loop integrand itself. We finish with a discussion of dual conformal symmetry in higher dimensions.

\section{On-shell superconformal symmetry}
\label{superconformalsymmetry}

An on-shell massless particle in four-dimensional Minkowski spacetime carries a light-like momentum $p$. The on-shell condition $p^2=0$ means that it is natural to express the momentum in terms of two commuting spinors,
\be
p^{\alpha \dot\alpha} = \lambda^\alpha \tilde{\lambda}^{\dot\alpha}\,.
\ee
In maximally supersymmetric Yang-Mills there are various species of on-shell particles which neatly organise into a superfield $\Phi$, dependent on Grassmann parameters $\eta^A$ which transform in the fundamental representation of the $su(4)$ R-symmetry. The on-shell superfield can be expanded as follows
\begin{equation}
\Phi = G^+ + \eta^A \Gamma_A + \tfrac{1}{2!} \eta^A \eta^B S_{AB} + \tfrac{1}{3!} \eta^A \eta^B \eta^C \epsilon_{ABCD} \overline{\Gamma}^D + \tfrac{1}{4!} \eta^A \eta^B \eta^C \eta^D \epsilon_{ABCD} G^-.
\label{onshellmultiplet}
\end{equation}
Here $G^+,\Gamma_A,S_{AB}=\tfrac{1}{2}\epsilon_{ABCD}\overline{S}^{CD},\overline{\Gamma}^A,G^-$ are the positive helicity gluon, gluino, scalar, anti-gluino and negative helicity gluon states respectively. $\mathcal{N}=4$ super Yang-Mills theory is a superconformal field theory and the on-shell superfield transforms under a natural action of the superconformal algebra which is essentially the oscillator representation \cite{Witten:2003nn}. Note that this same representation appears in the study of gauge-invariant local operators in the theory \cite{Beisert:2004ry} and plays a role in the integrable structure found in the spectrum of anomalous dimensions (see \cite{Beisert:2010jr} for a review). We give here the forms of the super Poincar\'e and superconformal generators,
\begin{align}
\label{superconformal}
& p^{\alpha \dot{\alpha} }  =   \lambda^{\alpha}  \tilde{\lambda}^{\dot{\alpha}}\,, & &
k_{\alpha \dot{\alpha}} =  \frac{\partial^2}{\partial \lambda^{\alpha} \partial \tilde{\lambda}^{ \dot{\alpha}}} \,,\notag\\
&q^{\alpha A} =  \lambda^{\alpha} \eta^A \,, &&   s_{\alpha A} =   \frac{\partial^2}{\partial \lambda^{\alpha} \partial \eta^{A}}\,, \notag\\
&  \bar{q}^{\dot\alpha}_A
= \tilde\lambda^{\dot \alpha} \frac{\partial}{\partial \eta^{A}}, & &
\bar{s}_{\dot\alpha}^A =  \eta^A \frac{\partial}{\partial \tilde{\lambda}^{\dot\alpha}}\,.
\end{align}
The Lorentz symmetry and $su(4)$ symmetry are manifest in this notation. The dilatation generator is given by
\be
d =   \frac{1}{2}\lambda^{\alpha} \frac{\partial}{\partial \lambda^{\alpha}} +\frac{1}{2} \tilde{\lambda}^{\dot{\alpha}} \frac{\partial}{\partial \tilde{\lambda}^{\dot{\alpha}}} +1.
\ee
There is also a central charge, also referred to as `helicity' or `little group weight',
\be
h = 1 + \frac{1}{2} \lambda^{\alpha} \frac{\partial}{\partial \lambda^{\alpha}} - \frac{1}{2} \tilde\lambda^{\dot \alpha} \frac{\partial}{\partial \tilde{\lambda}^{\dot \alpha}} - \frac{1}{2} \eta^A \frac{\partial}{\partial \eta^{A}}\,,
\ee
under which the superfield $\Phi$ has charge 1,
\begin{equation}
h \Phi = \Phi.
\end{equation}

When we consider superamplitudes, i.e. colour-ordered scattering amplitudes of on-shell superfields, then the helicity condition (or `homogeneity condition') is satisfied for each particle, 
\begin{equation}
h_i \mathcal{A}(\Phi_1,\ldots,\Phi_n) = \mathcal{A}(\Phi_1,\ldots,\Phi_n), \qquad i=1,\ldots,n.
\label{Ahelicity}
\end{equation}
The tree-level amplitudes in $\mathcal{N}=4$ super Yang-Mills theory can be written as follows,
\begin{equation}
\mathcal{A}(\Phi_1,\ldots,\Phi_n)=\mathcal{A}_n =  \frac{\delta^4(p) \delta^8(q)}{\langle12\rangle \ldots \langle n1\rangle } \mathcal{P}_n(\lambda_i,\tilde{\lambda}_i,\eta_i) = \mathcal{A}_n^{\rm MHV} \mathcal{P}_n.
\label{amp}
\end{equation}
The MHV tree-level amplitude,
\begin{equation}
\mathcal{A}_n^{\rm MHV} = \frac{\delta^4(p) \delta^8(q)}{\langle 12 \rangle \ldots \langle n1 \rangle},
\end{equation}
contains the delta functions $\delta^4(p) \delta^8(q)$ which are a consequence of translation invariance and supersymmetry and it can be factored out leaving behind a function with no helicity,
\begin{equation}
h_i \mathcal{P}_n = 0, \qquad i=1,\ldots,n.
\label{Phelicity}
\end{equation}
The function $\mathcal{P}_n$ can be expanded in terms of increasing Grassmann degree (the Grassmann degree always comes in multiples of 4 due to invariance under $su(4)$),
\begin{equation}
\mathcal{P}_n = 1 + \mathcal{P}_n^{\rm NMHV} + \mathcal{P}_n^{\rm NNMHV} + \,\, \ldots \,\,+ \mathcal{P}_n^{\overline{\rm MHV}}.
\end{equation}
The explicit form of the function $\mathcal{P}_n$ which encodes all tree-level amplitudes was found in \cite{Drummond:2008cr} by solving a supersymmetrised version \cite{Brandhuber:2008pf,ArkaniHamed:2008gz,Elvang:2008na} of the BCFW recursion relations \cite{Britto:2004ap,Britto:2005fq}.

At tree-level there are no infrared divergences and for generic configurations of the external momenta the amplitudes are annihilated by the generators of the standard superconformal symmetry, 
\begin{equation}
j_a \mathcal{A}_n = 0. \label{scs}
\end{equation}
Here we use the notation $j_a$ for any generator of the superconformal algebra $psu(2,2|4)$,
\begin{equation}
j_a \in \{p^{\alpha \dot \alpha},q^{\alpha A}, \bar{q}^{\dot \alpha}_A,m_{\alpha \beta}, \bar{m}_{\dot \alpha \dot \beta},r^A{}_B,d,s^\alpha_A,\bar{s}_{\dot \alpha}^A,k_{\alpha \dot \alpha} \}.
\end{equation}
For singular configurations there can be a non-zero variation \cite{Bargheer:2009qu,Korchemsky:2009hm,Sever:2009aa}). These variations can be systematically absorbed into a deformation of the generators acting on a generating functional for all tree-level amplitudes \cite{Bargheer:2009qu}. For more detail on these ideas see \cite{Bargheer:2011mm} in this review series.

\section{BCFW recursion relations}
\label{BCFW}

Let us consider gauge theory at tree-level. The Lagrangian contains three-gluon and four-gluon vertices. If we choose a standard Feynman gauge then the propagator assumes the form $1/p^2$. To calculate scattering amplitudes one should consider all amputated Feynman diagrams with a given choice of external legs, projected with the appropriate polarisation vectors to obtain the desired physical states. Stripping off the overall momentum-conserving delta-function, the end result of this process will be a rational function of the external momenta,
\be
\mathcal{A}_n = \delta^4(p) A_n\,.
\ee
All the poles of $A_n$ will be of the form
\be
\frac{i}{(p_i + \ldots + p_j)^2}\,
\ee
since they arise from internal propagators of the Feynman diagrams. 

In gauge theory one can also separate the amplitudes into various colour-ordered partial amplitudes, where the colour-dependence enters through a product of traces of the gauge group generators, carrying the colour labels of the scattering states (see \cite{Dixon:1996wi} or \cite{Dixon:2011xs} in this volume for more details). At tree-level it is sufficient to consider a single partial amplitude with colour structure ${\rm Tr} (T^{a_1} \ldots T^{a_n})$. All the others can then be obtained by non-cyclic permutations of the particle labels.

Knowing the behaviour of the amplitudes at the poles, it is possible to reconstruct the full function. This is the basic idea of BCFW recursion \cite{Britto:2005fq}. These ideas are reviewed in detail in \cite{Drummond:2010ep} and in this volume in \cite{Brandhuber:2011ke}. For our purposes here the important point is that one can reduce the problem to a single complex dimension by introducing a complex shift of the external momenta,
\begin{align}
\label{BCFWshift}
p_i^{\alpha \dot\alpha} \longrightarrow \hat{p}_i^{\alpha \dot\alpha}(z) &= (\lambda_i^\alpha - z \lambda_j^\alpha) \tilde{\lambda}_i^{\dot \alpha}\,, \notag \\
p_j^{\alpha \dot\alpha} \longrightarrow \hat{p}_j^{\alpha \dot \alpha}(z) &= \lambda_j^{\alpha} (\tilde{\lambda}_j^{\dot\alpha} + z \tilde{\lambda}_i^{\dot\alpha})\,.
\end{align}
Note that two legs ($i$ and $j$) have been selected to perform the shift.
Note that the external momenta remain on-shell (though complex) while at the same time preserving momentum-conservation. Those poles in the amplitude due to the propagators between legs $i$ and $j$ now manifest themselves as poles in $z$ in the shifted amplitude. The original unshifted amplitude can be reconstructed from a contour integral around the origin in the $z$-plane,
\be
A_n = A_n(0) = \oint \frac{dz}{2 \pi i z} A_n(z)\,.
\ee
On the other hand we can push the contour off to infinity. In doing so we will obtain a sum of residues from the poles in $A(z)$ and potentially also a contribution from infinity. Since the poles come from propagators in the Feynman diagrams we can interpret the residue as a product of two on-shell amplitudes on either side of the exchanged on-shell particle (summed over possible exchanged particles). The residues of the poles at finite $z$ are therefore related to amplitudes with a lower number of external legs. This is what allows a recursive construction of the tree-level amplitudes. In general the contribution from infinity only vanishes for certain choice of the helicities of the particles on the shifted legs. In the $\mathcal{N}=4$ theory which we are interested in here, all possible choices can be related to the $(++)$ shift by supersymmetry. Moreover it is natural to accompany the shift in the momenta (\ref{BCFWshift}) with a shift in the fermionic variables,
\be
q_j^{\alpha A} \longrightarrow \hat{q}_j^{\alpha A} = \lambda_j^\alpha(\eta_j^A + z \eta_i^A)\,.
\ee
It can be argued that this shift admits no contribution from infinity, for more details on this point see \cite{Britto:2005fq,ArkaniHamed:2008yf,Brandhuber:2008pf,ArkaniHamed:2008gz}. Thus the amplitudes can be constructed entirely from lower-point data.

\begin{figure}
\psfrag{hat1}[cc][cc]{$\hat{1}$}
\psfrag{hatn}[cc][cc]{$\hat{n}$}
\psfrag{AL}[cc][cc]{\,\,$A_L$}
\psfrag{AR}[cc][cc]{\,\,\,$A_R$}
\psfrag{i-1}[cc][cc]{$i-1$}
\psfrag{i}[cc][cc]{$i$}
\psfrag{s}[cc][cc]{$s$}
\psfrag{sb}[cc][cc]{$\bar{s}$}
\psfrag{sum}[cc][cc]{$\sum_{s}$}
 \centerline{{\epsfysize5cm
\epsfbox{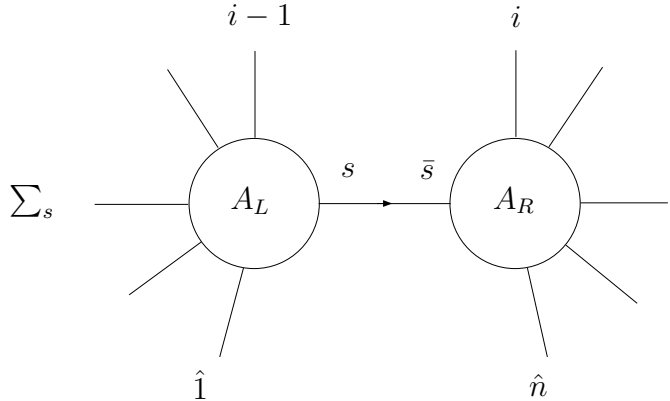}}}  \caption[]{\small The residues at the poles of $A_n(z)$ factorise into products of lower-point amplitudes. We denote the location of pole by $z_{P_i}$ where $P_i$ is the momentum in that channel of the unshifted diagram $P_i = p_1 + \ldots p_{i-1}$. For each pole there is a sum over possible exchanged states $s$ as we consider all possible Feynamn diagrams with the relevant propagator. In the maximally supersymmetric theory this sum is captured by a Grassmann integral.}
  \label{statesum}
\end{figure}

As we have seen the states in $\mathcal{N}=4$ super Yang-Mills theory are arranged into a single on-shell supermultiplet. The sum over possible exchanged states can be replaced by a single Grassmann integral over the $\eta$ variable associated to the internal line joining the two subamplitudes in the recursion relation. In summary the recursion relation for $\mathcal{N}=4$ super Yang-Mills theory is
\be
A_n = \sum_i \int  d^4 \eta_{\hat{P}_i} A_L\bigl( \hat{1}(z_{P_i}),2,\ldots,i-1,-\hat{P}(z_{P_i})\bigr) \biggl(\frac{i}{P_i^2}\biggr)  A_R \bigl( \hat{P}(z_{P_i}),i,\ldots,n-1,\hat{n}(z_{P_i}) \bigr)\,.
\label{superBCFW}
\ee
Here the sum is over the different residues obtained by dragging the contour off to infinity in the complex $z$-plane. For each residue $z_{P_i}$ is the associated location of the pole, while $P_i$ is the unshifted momentum, $P_i = p_1 + \ldots p_{i-1}$ in the associated channel. Finally $\hat{P}(z_{P_i})$ is the shifted quantity $\hat{P} = \hat{p}_1 +p_2 + \ldots p_{i-1}$ evaluated at the pole value of $z$ so that it becomes an on-shell momentum. Note that to simplify the formula, we have chosen to perform the shift for momenta $p_1$ and $p_n$.

In order to solve the recursion relation (\ref{superBCFW}) we need on-shell three-point amplitudes, which can arise from Feynman diagrams with three-point vertices attached to two external legs.
The fact that the momenta are complex is important because for real momenta the three-point amplitude vanishes. There are two solutions to the on-shell three-point kinematics. From momentum conservation,
\be
\lam_1^\a \tlam_1^\adt + \lam_2^\a \tlam_2^\adt + \lam_3 \tlam_3^\adt = 0 
\label{3ptmomcons}
\ee
we find the two possibilities
\begin{align}
&\l< 12 \r> = \l< 23 \r> = \l<31 \r> = 0 \qquad (\overline{\rm MHV}), \\
&[12] \,=\, [23] \, = \, [31] \,=0 \qquad \text{(MHV)}.
\end{align}
The corresponding superamplitudes are fixed by helicity, Lorentz invariance, momentum conservation and supersymmetry \cite{Brandhuber:2008pf,ArkaniHamed:2008gz},
\be
\mathcal{A}_3^{\rm MHV} = \frac{\delta^4(p) \delta^8(q)}{\l< 12 \r> \l<23 \r> \l< 31 \r>}.
\label{3ptMHV}
\ee
\be
\mathcal{A}_3^{\overline{\rm MHV}} = \frac{\delta^4(p) \delta^4(\eta_1 [23] + \eta_2 [31] + \eta_3 [12])}{[12][23][31]}.
\label{3ptMHVbar}
\ee
Note that in the three-point $\overline{\rm MHV}$ kinematics the supersymmetry generator factorises, $q^{\a A} = \lam_F^\a q_F^A$ for some $\lam_F^\a$ and $q_F^A$ and the requirement of $q$-supersymmetry is only that the amplitude contain a factor of $\delta^4(q_F)$ and not $\delta^8(q)$. Thus the $\overline{\rm MHV}_3$ amplitude has Grassmann degree four while all other amplitudes have Grassmann degree at least eight. 

The recursion relation (\ref{superBCFW}) can be decomposed into contributions of various Grassmann degrees. Since there is a Grassmann integral on the RHS the sum of the degrees of the two subamplitudes $A_L$ and $A_R$ must be four more than the degree of the amplitude we are solving for. Thus we find
\begin{eqnarray}
\label{super-BCF-all} 
A_{n}^{{\rm
N}^{p}{\rm MHV}} &=& 
\int \frac{d^{4}
\eta_{\hat{P}}}{P^2} \, A_{3}^{\rm \widebar{MHV}}(z_{P})
A_{n-1}^{{\rm N}^{p}{\rm MHV}}(z_{P}) 
\nonumber \\
 &+& 
 \sum_{m=0}^{p-1} \; \; \sum_i  
\int \frac{d^{4}\eta_{\hat{P}_{i}}}{P_i^2}
A_{i}^{{\rm N}^{m}{\rm MHV}}(z_{P_{i}})
A_{n-i+2}^{{\rm N}^{(p-m-1)}{\rm MHV}}(z_{P_{i}})\,.
\end{eqnarray}
Note that we have not allowed for the left subamplitude to be $A^{\rm MHV}_3$. This is consistent because in the MHV case the square bracket $[12]$ vanishes. For the left subamplitude the $\tlam$ variables are unshifted and hence this would imply that $[12]$ and hence $(p_1+p_2)^2$ vanishes for the full amplitude as well. This is a restriction on the kinematics which is not true in general and hence such a term does not contribute to the recursion relation. Similarly the right subamplitude can never be $A^{\overline{\rm MHV}}_3$.

\begin{figure}
\psfrag{hat1}[cc][cc]{$\hat{1}$}
\psfrag{hat4}[cc][cc]{$\hat{n}$}
\psfrag{2}[cc][cc]{$2$}
\psfrag{3}[cc][cc]{$3$}
\psfrag{MHVb}[cc][cc]{$\overline{\rm MHV}$}
\psfrag{MHV}[cc][cc]{MHV}
\psfrag{P}[cc][cc]{$\hat{P}$}
 \centerline{{\epsfysize4cm
\epsfbox{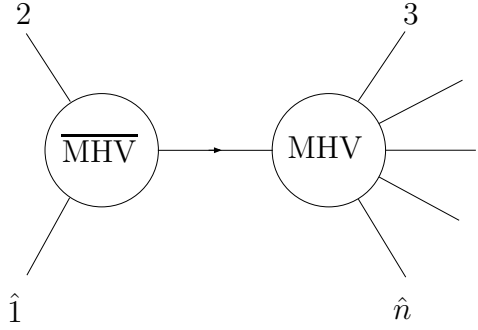}}}  \caption[]{\small The single BCFW diagram contributing to the $n$-point MHV amplitude.}
  \label{nptamp}
\end{figure}
The simplest case of the recursion relation is for the MHV amplitudes. In this case the relation reduces to
\be
A_{n}^{{\rm MHV}} = 
\int \frac{d^{4}
\eta_{\hat{P}}}{P^2} \, A_{3}^{\rm \widebar{MHV}}(z_{P})
A_{n-1}^{{\rm MHV}}(z_{P}) \,.
\ee
The single term on the RHS is illustrated in Fig. \ref{nptamp}.
The solution to the MHV equation \cite{Brandhuber:2008pf} is the natural generalisation of (\ref{3ptMHV}) to $n$ points,
\be
A_n^{\rm MHV} = \frac{\delta^4(q)}{\langle 12 \rangle \ldots \langle n1\rangle}\,.
\ee
It is the superspace formula for the MHV amplitudes of Nair \cite{Nair:1988bq}.

\begin{figure}
\psfrag{dots}[cc][cc]{$\ldots$}
\psfrag{hat1}[cc][cc]{$\hat{1}$}
\psfrag{2}[cc][cc]{$2$}
\psfrag{3}[cc][cc]{$3$}
\psfrag{hatn}[cc][cc]{$\hat{n}$}
\psfrag{i-1}[cc][cc]{$i-1$}
\psfrag{P}[cc][cc]{$\hat{P}$}
\psfrag{Pi}[cc][cc]{$\hat{P}_{i}$}
\psfrag{i}[cc][cc]{\hspace{0cm}$i$}
\psfrag{MHV}[cc][cc]{MHV}
\psfrag{NMHV}[cc][cc]{NMHV}
\psfrag{MHVb}[cc][cc]{$\overline{\rm MHV}$}
\psfrag{+sum}[cc][cc]{$\,\,\,\,\,\, + \,\,\,\,\,\, \sum_{i=4}^{n-1}$}
 \centerline{{\epsfysize4cm
\epsfbox{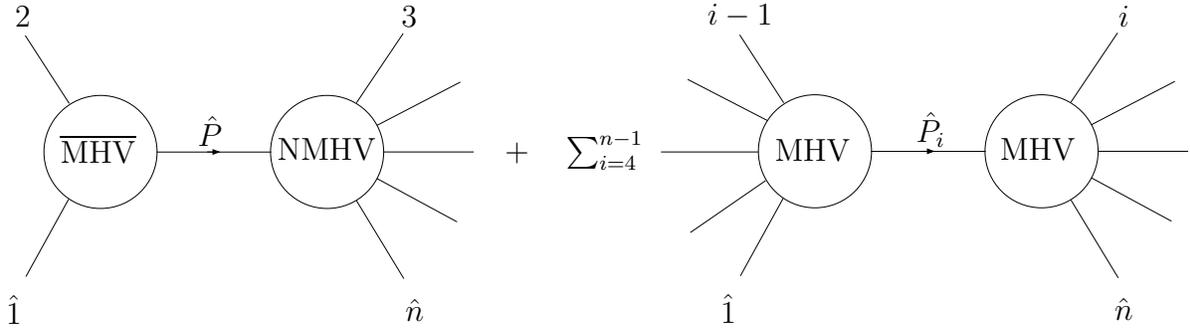}}}  \caption[]{\small
  The two contributions to the supersymmetric recursion relation for
  NMHV amplitudes. We call them homogeneous and
  inhomogeneous respectively.}
  \label{fig1}
\end{figure}

Let us now consider the next-to-MHV (NMHV) case. The recursion relation has two types of terms, one where the subamplitudes are $A^{\overline{\rm MHV}}_3$ and $A_{n-1}^{\rm NMHV}$ and the other where they are both MHV,
\be
A_n^{\rm NMHV} = \int \frac{d^{4}
\eta_{\hat{P}}}{P^2} \, A_{3}^{\rm \widebar{MHV}}(z_{P})
A_{n-1}^{{\rm N}^{p}{\rm MHV}}(z_{P}) + \sum_{i=3}^{n-1} \int \frac{d^{4}\eta_{\hat{P}_{i}}}{P_i^2}
A_{i}^{{\rm MHV}}(z_{P_{i}})
A_{n-i+2}^{{\rm MHV}}(z_{P_{i}})\,.
\label{NMHVBCFW}
\ee
The two kinds of terms are represented in Fig. \ref{fig1}.

To express the solution for the NMHV amplitudes it is very convenient to introduce dual superspace coordinates. This notation is directly connected to the existence of a new `dual' superconformal symmetry of the amplitudes which we will discuss in the next section. The dual coordinates will be defined in terms of the momenta and supercharges,
\be
p_i^{\alpha \dot\alpha} =\lambda_i^\alpha \tilde{\lambda}_i^{\dot \alpha}= x_i^{\alpha \dot\alpha} - x_{i+1}^{\alpha \dot\alpha}\, \qquad q_i^{\alpha A} = \lambda_i^{\alpha} \eta_i^A = \theta_i^{\alpha A} - \theta_{i+1}^{\alpha A}\,.
\label{dualcoords}
\ee
The solution to the NMHV recursion relation (\ref{NMHVBCFW}) is then \cite{Drummond:2008cr}
\be
A_n^{\rm NMHV} = A_n^{\rm MHV} \mathcal{P}_n^{\rm NMHV},
\label{ANMHV}
\ee
where
\be
\mathcal{P}_n^{\rm NMHV} = \sum_{2\leq a<b\leq n-1} \!\!\!\!\!\! R_{n,ab}
\label{PNMHV}
\ee
and
\be
R_{n,ab} = \frac{\l< a \, a-1\r> \l< b \, b-1\r> \delta^4 \bigl(\l< n |x_{na}x_{ab}|\theta_{bn}\r> + \l<n|x_{nb}x_{ba}|\theta_{an}\r>\bigr)}{x_{ab}^2 \l< n |x_{na} x_{ab} |b\r> \l< n|x_{na} x_{ab} |b-1\r> \l<n|x_{nb}x_{ba}|a\r> \l<n|x_{nb}x_{ba}|a\r> \l<n|x_{nb}x_{ba}|a-1\r>}
\label{Rnab}
\ee
and the sum over $a$ and $b$ in (\ref{PNMHV}) is performed such that $a<b-1$. The solution for the NMHV amplitudes coincides with that found in \cite{Drummond:2008vq} as a supersymmetrisation of the three-mass box coefficients of \cite{Bern:2004bt}.
The final result (\ref{PNMHV}) is remarkably simple. The six-particle case, for example, is expressed as sum of only three terms.

\begin{figure}
\psfrag{one}[cc][cc]{$1$}
\psfrag{a1b1}[cc][cc]{$a_{1}b_{1}$}
\psfrag{a2b2}[cc][cc]{$a_{2}b_{2}$}
\psfrag{a3b3}[cc][cc]{$a_{3}b_{3}$}
\psfrag{b1a1a2b2}[cc][cc]{$b_{1}a_{1};a_{2}b_{2}$}
\psfrag{b2a2a3b3}[cc][cc]{$b_{2}a_{2};a_{3}b_{3}$}
\psfrag{b1a1a3b3}[cc][cc]{$b_{1}a_{1};a_{3}b_{3}$}
\psfrag{b1a1b2a2a3b3}[cc][cc]{$b_{1}a_{1};b_{2}a_{2};a_{3}b_{3}$}
\psfrag{two}[cc][cc]{$2$}
\psfrag{n1}[cc][cc]{$n-1$}
\psfrag{a1p}[cc][cc]{$a_{1}+1$}
\psfrag{a2p}[cc][cc]{$a_{2}+1$}
\psfrag{b1}[cc][cc]{$b_{1}$}
\psfrag{b2}[cc][cc]{$b_{2}$}
 \centerline{{\epsfysize7cm
\epsfbox{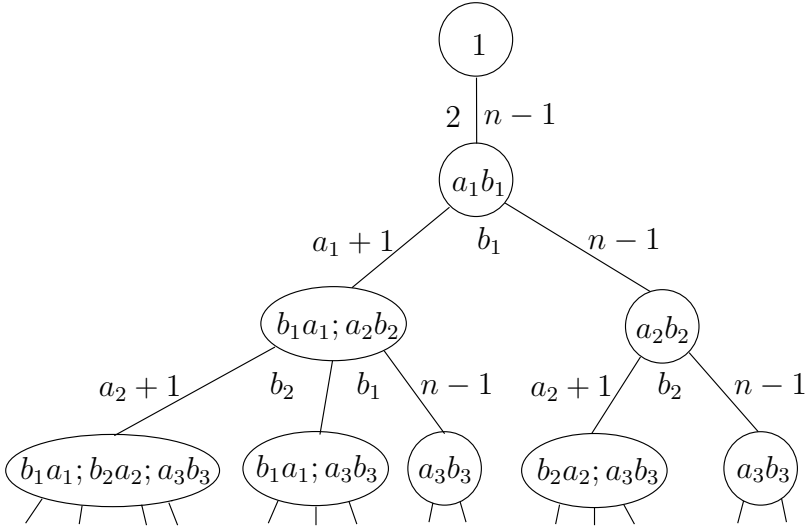}}}  \caption[]{\small
  Graphical representation of the formula for tree-level amplitudes in
  $\mathcal{N}=4$ SYM.}
  \label{fig-rec-solution}
\end{figure}

The process of solving the recursion relation can be continued to higher levels in the MHV degree. Instead of working level by level however, it is possible to write down an expression which can be verified to satisfy the recursion relation (\ref{superBCFW}) directly \cite{Drummond:2008cr}. It is helpful to strip off the MHV prefactors for the superamplitudes and just work in terms of the function $\mathcal{P}_n$. Having done this the recursion relation can be expressed as
\be
\mathcal{P}_n = \mathcal{P}_{n-1}(\hat{P},3,\ldots,\hat{n}) + \sum_{i=4}^{n-1} R_{n;2,i} \mathcal{P}_i(2,\ldots,-\hat{P_i},\hat{1}) \mathcal{P}_{n-i+2}(\hat{P_i},i,\ldots,\hat{n}).
\label{fullPrec}
\ee
Note that the same quantity $R_{n,ab}$ appears in the quadratic term on the RHS. The solution to this equation is encoded diagrammatically in Fig. \ref{fig-rec-solution}. To obtain the expression for the tree-level amplitudes from Fig. \ref{fig-rec-solution} we consider paths, starting at the top and descending through the tree structure. Each node of the path corresponds to a generalisation of formula (\ref{Rnab}) carrying the labels of that node,
\be
R_{n;b_1a_1;b_2a_2;\ldots;b_ra_r;ab} = \frac{\l< a\,\,a-1\r> \l<
  b\,\,b-1\r> \delta^{(4)}(\l< \xi | x_{a_r a}x_{ab} | \theta_{ba_r} \r> +
  \l< \xi |x_{a_r b} x_{ba} | \theta_{aa_r} \r>)}{x_{ab}^2 \l< \xi |
  x_{a_r a} x_{ab} | b \r> \l< \xi | x_{a_r a} x_{ab} |b-1\r> \l< \xi
  | x_{a_r b} x_{ba} |a\r> \l< \xi | x_{a_r b} x_{ba} |a-1\r>}\,,
  \label{generalR}
\ee
with
\be
\l< \xi | = \l<n | x_{nb_1}x_{b_1 a_1} x_{a_1 b_2} x_{b_2 a_2} \ldots x_{b_r a_r} \,.
\ee
For a given path we are required to take the product over all nodes and for each node perform a summation over the final pair of labels over the region $L\leq a < b \leq U$  where the limits $L,U$ are given on either side of the line above each node in Fig. \ref{fig-rec-solution}. A final subtlety is that when the labels take their boundary values, formula (\ref{generalR}) must be modified slightly so that the explicit dependence on the boundary spinor is replaced. This is dealt with by writing superscripts on the $R$-invariants to indicate special behaviour for terms when $a=L$ or $b=U$.
Specifically we write
\be
\sum_{L\leq a<b \leq U} \!\!\!\!\!\! R_{n;b_1a_1;\ldots;b_ra_r;ab}^{l_1\ldots l_p;u_1\ldots u_q} \, .
\label{superscripts}
\ee
This notation means that for the terms in the sum where $a=L$ (or $b=U$) we replace the explicit dependence on $\langle L-1|$ (or $\langle U|$) in (\ref{generalR}) in the following way,
\be
\langle L-1 | \longrightarrow \langle n|x_{nl_1} x_{l_1l_2} x_{l_2l_3} \ldots x_{l_{p-1} l_p} \,, \qquad \langle U | \longrightarrow \langle n | x_{nu_1} x_{u_1u_2} x_{u_2u_3} \ldots x_{u_{q-1}u_q} \, .
\label{Lrep}
\ee
When no replacement is to be made on one of the boundaries we will write the superscript $0$. The right superscripts are obtained by taking the lables of the node, deleting the final pair and flipping the order of the preceding pair. The left superscripts are identified with the right superscripts of the node to the left in a given row of the tree diagram Fig. \ref{fig-rec-solution}.

As an example we will give the result for NNMHV amplitudes,
\begin{align}
\label{PNNMHVnew}
\mathcal{P}_n^{\rm NNMHV} = \sum_{2\leq a_1,b_1 \leq n-1}
\!\!\!\!\!\!\!\! R_{n;a_1b_1} \Biggl[ &\sum_{a_1+1 \leq a_2,b_2 \leq
  b_1} \!\!\!\!\!\!\!\! R_{n;b_1a_1;a_2b_2}^{0;a_1b_1}
  +  \sum_{b_1 \leq a_2b_2 \leq n-1} \!\!\!\!\!\!\!\! R_{n;a_2b_2}^{a_1b_1;0} \Biggr]\,.
\end{align}

At the level of pure gluon scattering, the fact that we have solved for the amplitudes in the $\mathcal{N}=4$ theory is no restriction at all; the gluon amplitudes are the same in any gauge theory. Thus the simplicity of the expressions arising from the recursive structure is universal for gluon amplitudes in all gauge theories, as is the associated presence of spurious non-local poles. The explicit expressions for the pure-gluon amplitudes can be derived from the superamplitudes by reading off the coefficient of the relevant combination of $\eta$ variables. Some specific examples are described in \cite{Drummond:2008cr}, in particular one can check that the expressions for the split-helicity amplitudes agree with those computed in \cite{Britto:2005dg}. The results for amplitudes involving fermions can be adapted to other gauge theories, for example QCD \cite{Dixon:2010ik}. 

The same recursive technique is also valid for gravitational theories \cite{Bedford:2005yy,Cachazo:2005ca,ArkaniHamed:2008yf}. Again it can be made manifestly supersymmetric and becomes much simpler for the maximally supersymmetric theory $\mathcal{N}=8$ supergravity \cite{ArkaniHamed:2008gz}, admitting an explicit solution \cite{Drummond:2009ge}. 


\section{Dual superconformal symmetry}
\label{dualsconf}

Now let us look at the expressions we have obtained for the scattering amplitudes arising from the solution of the BCFW recursion relation from the point of view of symmetry. The quantities $x_i$ which we introduced in (\ref{dualcoords}) to express the amplitudes can be interpreted as the coordinates of a dual copy of spacetime. The amplitudes are trivially invariant under translations of the dual coordinates as they were introduced only through their differences. The definition of the dual coordinates is covariant and so dual Lorentz transformations are the same as Lorentz transformations of the particle momenta and so are also a symmetry of the scattering amplitudes. 
Perhaps surprisingly, it turns out that conformal transformations of the $x_i$ are also a  symmetry of the scattering amplitudes. 
Since the symmetry acts canonically on the dual coordinates and these are linearly related to the particle momenta the generator of the transformation is a first-order operator acting on the momenta. Note that such a conformal transformation is not related to the conformal symmetry of the Lagrangian, which instead is related to the second-order generators $k_{\a \adt}$ of (\ref{superconformal}). The conformal symmetry acting in the dual space is referred to as dual conformal symmetry. We will now see explicitly how the forms for the tree-level amplitudes in the previous section reveal the new symmetry.

The dual coordinates $x_i$ define a closed polygon with light-like edges in the dual space, as represented in Fig. 11.
The contour is closed because we identify $x_{n+1}$ with $x_1$. This statement reflects the total momentum conservation of the scattering process $p_1+p_2+\ldots +p_n =0$. The edges of the polygon are light-like because the particles in the scattering process are on-shell and the ordering of the edges is given by the order of the particles in the overall colour trace of the ordered partial amplitude. The role of the polygon was first made clear at strong coupling \cite{Alday:2007hr} where it becomes the boundary for a minimal surface in AdS space, leading to the idea of a relation between scattering amplitudes and Wilson loops \cite{Drummond:2007aua,Brandhuber:2007yx,Drummond:2007cf,Drummond:2007au,Drummond:2007bm,Bern:2008ap,Drummond:2008aq,Anastasiou:2009kna}. This relation is reviewed in more detail in \cite{Drummond:2010km}.
For now we would just like to
note that a light-like polygon maps into another such polygon under conformal transformations of the dual space. Indeed under conformal inversions
\be
x^\mu \longrightarrow  \frac{x^\mu}{x^2},
\ee
we have that 
\be
x_{ij}^2 \longrightarrow \frac{x_{ij}^2}{x_i^2 x_j^2}.
\ee
Thus if two points $x_i$ and $x_j$ are light-like separated they will remain so after a conformal inversion. The conformal group is generated by Lorentz transformations, translations and conformal inversions so the light-like nature of the polygon is invariant under the action of the whole conformal group. 

\begin{figure}
\psfrag{x1}[cc][cc]{$x_1$}
\psfrag{x2}[cc][cc]{$x_2$}
\psfrag{x3}[cc][cc]{$x_3$}
\psfrag{x4}[cc][cc]{$x_4$}
\psfrag{xn}[cc][cc]{$x_n$}
\psfrag{p1}[cc][cc]{$p_1$}
\psfrag{p2}[cc][cc]{$p_2$}
\psfrag{p3}[cc][cc]{$p_3$}
\psfrag{pn}[cc][cc]{$p_n$}
 \centerline{{\epsfysize6cm
\epsfbox{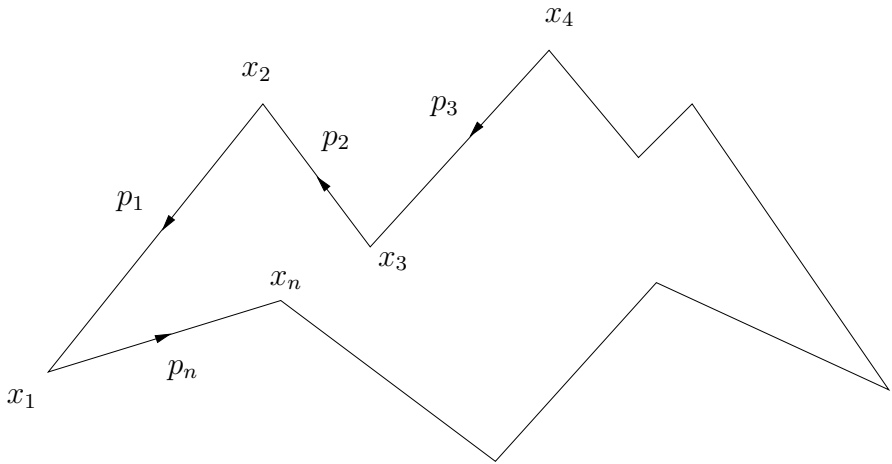}}}  \caption[]{\small The light-like polygon in dual coordinate space defined by the particle momenta.}
  \label{polygon}
\end{figure}

To see that dual conformal transformations are actually a symmetry of the scattering amplitudes we need to define their action on all of the variables in the problem. 
The helicity variables $\lam$ and $\tlam$ must also transform under dual conformal transformations as they are related to the dual coordinates via the defining relations,
\be
x_i^{\a \adt} - x_{i+1}^{\a \adt} = \lam_i^\a \tlam_i^{\adt}.
\label{dualx}
\ee
Under a conformal inversion $x_{ij}^{\a \adt}$ transforms as follows,
\be
x_{ij}^{\a \adt} \longrightarrow  -( x_i^{-1}x_{ij}x_j^{-1} )^{\a \adt}.
\ee
In order to ensure compatibility with this we can choose the transformations $\lam_i$ and $\tlam_i$ to be given by \cite{Drummond:2008vq},
\be
\lam_i^{\a} \longrightarrow (x_i^{-1} \lam_i)^{\adt}, \qquad \tlam_i^{\adt} \longrightarrow -(x_{i+1}^{-1} \tlam_i)^{\a}.
\ee
Note that we could have distributed the weights between $\lambda_i$ and $\tilde\lambda_i$ differently and still maintained compatibility with (\ref{dualx}). The choice made above means that the $\lambda_i$ transform in a similar way to the odd dual coordinates,
\be
\theta_i^{\a A} \longrightarrow (x_i^{-1}\theta_i)^{\adt A}.
\ee
The above transformation for $\theta_i$ implies that the variables $\eta_i$ must also transform in analogy with $\lambda_i$ and $\tlam_i$, although we will not need the explicit variation here.

Now let us examine the explicit forms of the tree-level amplitudes.
If we look at the MHV amplitude,
\be
\mathcal{A}_n^{\rm MHV} = \frac{\delta^4(p) \delta^8(q)}{\l< 12 \r> \ldots \l<n 1 \r>},
\ee 
we can see that it is in fact covariant under dual conformal transformations. Firstly, if we drop the requirement that $x_{n+1} \equiv x_1$ and $\theta_{n+1} \equiv \theta_1$ then the delta functions can be written as
\be
\delta^4(p) \delta^4(q) = \delta^4(x_1 - x_{n+1}) \delta^8(\theta_1 - \theta_{n+1}).
\label{deltas}
\ee 
This combination is dual conformally invariant since the bosonic delta function has conformal weight 4 at the point $x_1$ (which is identified with $x_{n+1}$ under the delta function) as can be seen from $\int d^4 x_1 \delta^4(x_1 - x_{n+1}) =1$. The Grassmann delta function has the opposite conformal weight (which is only true because of maximal supersymmetry) and the hence the product (\ref{deltas}) is invariant.

The denominator in the MHV amplitude is covariant under dual conformal transformations because factors of the form $\l< i \,i+1\r>$ transform as follows,
\be
\l<i \, i+1\r> \longrightarrow \l<i|x_i^{-1} x_{i+1}^{-1} |i+1\r> = \frac{\l< i|x_i x_{i+1} |i+1\r>}{x_i^2 x_{i+1}^2} = \frac{\l< i |x_{i+1} x_{i+1} |i+1 \r>}{x_i^2 x_{i+1}^2} = \frac{\l<i \, i+1\r>}{x_i^2}.
\label{abcovariance}
\ee
Thus we find that the MHV tree-level amplitude is covariant with weight 1 at each point,
\be
\mathcal{A}_n^{\rm MHV} \longrightarrow (x_1^2 \ldots x_n^2)\mathcal{A}_n^{\rm MHV}.
\ee

If we now look at the NMHV amplitude defined by equations (\ref{ANMHV}), (\ref{PNMHV}) and (\ref{Rnab}) we find that it is similarly covariant. The reason is that each term $R_{n,ab}$ in $\mathcal{P}_n^{\rm NMHV}$ is invariant under dual conformal transformations. Indeed returning to the formula (\ref{Rnab}) we see that it is made of dual conformally covariant factors. The spurious poles are covariant following a similar analysis to (\ref{abcovariance}).  For example we have
\be
\l< n|x_{na} x_{ab} |b \r> \longrightarrow \frac{\l<n|x_{na}x_{ab}|b\r>}{x_n^2 x_a^2 x_b^2}, \qquad \l<n|x_{na}x_{ab} |b-1\r> \longrightarrow \frac{\l<n| x_{na}x_{ab}|b-1\r>}{x_n^2 x_a^2 x_{b-1}^2}.
\ee
The Grassmann delta function is also covariant as can be see when written in a slightly different form,
\be
\delta^4 \bigl(\l< n |x_{na}x_{ab}|\theta_{bn}\r> + \l<n|x_{nb}x_{ba}|\theta_{an}\r>\bigr) = \delta^4 \bigl(\l< n |x_{na}x_{ab}|\theta_{b}\r> + \l<n|x_{nb}x_{ba}|\theta_{a}\r> + x_{ab}^2 \l< n \theta_n \r>\bigr)\,.
\ee
Checking all the factors in (\ref{Rnab}) one can see that the weights cancel and thus $R_{n,ab}$ is invariant under dual conformal transformations. In fact one can show from the recursion relation itself that all terms produced this way will respect dual conformal symmetry \cite{Brandhuber:2008pf}. In the language of the previous section one can see this fact in (\ref{fullPrec}). If we suppose that $\mathcal{P}_k$ is invariant for all $k$ up to $n-1$ then all quantities appearing on the RHS of (\ref{fullPrec}) are invariants under dual conformal symmetry and hence so is $\mathcal{P}_n$.
One can also see directly, using transformation rules similar to (\ref{abcovariance}), that dual conformal symmetry is present for the generalisations of $R_{n,ab}$ appearing in the N$^p$MHV amplitudes.

It is perhaps not surprising, given that we are dealing with a supersymmetric theory, that the dual conformal symmetry extends to a dual superconformal one \cite{Drummond:2008vq}.
Dual superconformal symmetry has a canonical action on the coordinates of the dual superspace $x_i,\theta_i$.
For many purposes it is helpful to express the symmetry in terms of the generators of infinitesimal transformations. For example the generator of special conformal transformations of the dual coordinates is
\be
K_{\alpha \dot{\alpha}} = \sum_i [x_{i \alpha}{}^{\dot{\beta}} x_{i
    \dot{\alpha}}{}^{\beta} \partial_{i \beta \dot{\beta}} + x_{i
    \dot{\alpha}}{}^{\beta} \theta_{i \alpha}^B \partial_{i \beta B}].
\label{dualK}
\ee
Just as we have seen for dual conformal inversions the transformation must act on the on-shell superspace variables $\{\lam,\tlam,\eta\}$ in order to respect the constraints between them. In terms of the generators this means we must add terms so that the generator commutes with the constraints modulo constraints. One can perform this process for all generators of the superconformal algebra. The result is the following set of generators \cite{Drummond:2008vq},
\begin{align}
P_{\alpha \dot{\alpha}}&= \sum_i \partial_{i \alpha \dot{\alpha}}\,, \qquad Q_{\alpha A} = \sum_i \partial_{i \alpha A}\,, \qquad
\overline{Q}_{\dot{\alpha}}^A = \sum_i [\theta_i^{\alpha A}
  \partial_{i \alpha \dot{\alpha}} + \eta_i^A \partial_{i \dot{\alpha}}], \notag\\
M_{\alpha \beta} &= \sum_i[x_{i ( \alpha}{}^{\dot{\alpha}}
  \partial_{i \beta ) \dot{\alpha}} + \theta_{i (\alpha}^A \partial_{i
  \beta) A} + \lambda_{i (\alpha} \partial_{i \beta)}]\,, \qquad
\overline{M}_{\dot{\alpha} \dot{\beta}} = \sum_i [x_{i
    (\dot{\alpha}}{}^{\alpha} \partial_{i \dot{\beta} ) \alpha} +
  \tilde{\lambda}_{i(\dot{\alpha}} \partial_{i \dot{\beta})}]\,,\notag\\
R^{A}{}_{B} &= \sum_i [\theta_i^{\alpha A} \partial_{i \alpha B} +
  \eta_i^A \partial_{i B} - \tfrac{1}{4} \delta^A_B \theta_i^{\alpha
    C} \partial_{i \alpha C} - \tfrac{1}{4}\delta^A_B \eta_i^C \partial_{i C}
]\,,\notag\\
D &= \sum_i [-x_i^{\dot{\alpha}\alpha}\partial_{i \alpha \dot{\alpha}} -
  \tfrac{1}{2} \theta_i^{\alpha A} \partial_{i \alpha A} -
  \tfrac{1}{2} \lambda_i^{\alpha} \partial_{i \alpha} -\tfrac{1}{2}
  \tilde{\lambda}_i^{\dot{\alpha}} \partial_{i \dot{\alpha}}]\,,\notag\\
C &=  \sum_i [-\tfrac{1}{2}\lambda_i^{\alpha} \partial_{i \alpha} +
  \tfrac{1}{2}\tilde{\lambda}_i^{\dot{\alpha}} \partial_{i \dot{\alpha}} + \tfrac{1}{2}\eta_i^A
  \partial_{i A}]\,, \notag\\
S_{\alpha}^A &= \sum_i [-\theta_{i \alpha}^{B} \theta_i^{\beta A}
  \partial_{i \beta B} + x_{i \alpha}{}^{\dot{\beta}} \theta_i^{\beta
    A} \partial_{i \beta \dot{\beta}} + \lambda_{i \alpha}
  \theta_{i}^{\gamma A} \partial_{i \gamma} + x_{i+1\,
    \alpha}{}^{\dot{\beta}} \eta_i^A \partial_{i \dot{\beta}} -
  \theta_{i+1\, \alpha}^B \eta_i^A \partial_{i B}]\,,\notag\\
\overline{S}_{\dot{\alpha} A} &= \sum_i [x_{i \dot{\alpha}}{}^{\beta}
  \partial_{i \beta A} + \tilde{\lambda}_{i \dot{\alpha}}
  \partial_{iA}]\,,\notag\\
K_{\alpha \dot{\alpha}} &= \sum_i [x_{i \alpha}{}^{\dot{\beta}} x_{i
    \dot{\alpha}}{}^{\beta} \partial_{i \beta \dot{\beta}} + x_{i
    \dot{\alpha}}{}^{\beta} \theta_{i \alpha}^B \partial_{i \beta B} +
  x_{i \dot{\alpha}}{}^{\beta} \lambda_{i \alpha} \partial_{i \beta}
  + x_{i+1 \,\alpha}{}^{\dot{\beta}} \tilde{\lambda}_{i \dot{\alpha}}
  \partial_{i \dot{\beta}} + \tilde{\lambda}_{i \dot{\alpha}} \theta_{i+1\,
    \alpha}^B \partial_{i B}]\,.
\label{dualsc}
\end{align}
Here we have employed the following shorthand notation
\begin{align}\label{shortderiv}
\partial_{i \alpha \dot{\alpha}} = \frac{\partial}{\partial
x_i^{\alpha \dot{\alpha}}}, \qquad \partial_{i \alpha A} = \frac{\partial}{\partial \theta_i^{\alpha
A}}, \qquad \partial_{i \alpha} = \frac{\partial}{\partial \lambda_i^{\alpha}}\,, \qquad
\partial_{i \dot{\alpha}} = \frac{\partial}{\partial
    \tilde{\lambda}_i^{\dot{\alpha}}}\,, \qquad
\partial_{i A} = \frac{\partial}{\partial \eta_i^A}\,.
\end{align}

It is simple to check that the generators in (\ref{dualsc}) do obey the commutation relations of the superconformal algebra. There are several remarks worth making at this point. Firstly it is clear from the first-order form of the generators that the dual superconformal symmetry is distinct from the ordinary superconformal symmetry generated by the operators given in (\ref{superconformal}). Secondly, we note that the $su(4)$ nature of the fermionic generators is swapped between the original superconformal algebra and the dual superconformal algebra. For example the supersymmetry generator $q^{\a A}$ is in the fundamental representation of $su(4)$ while the dual supersymmetry generator $Q_{\a A}$ is in the anti-fundamental. Similarly, on the on-shell superspace variables, the two dilatation generators coincide up to a sign because dual coordinates are actually related to particle momenta. Finally we note that the two superconformal algebras overlap non-trivially. That is, the fermionic superconformal generator $\bar{S}$ coincides with the original supersymmetry generator $\bar{q}$ on the on-shell superspace, while the dual supersymmetry generator $\bar{Q}$ coincides with the original superconformal generator $\bar{s}$. The definitions of the dual variables manifestly respect covariance under the Lorentz, dilatation and $su(4)$ symmetries and so these symmetries are shared between the two copies of the superconformal algebra. 
The same structure also arises from considering the symmetries of the string sigma model \cite{Ricci:2007eq,Berkovits:2008ic,Beisert:2008iq} which is believed to describe the theory at strong coupling.

With all the generators of the superconformal algebra to hand we can now verify that the quantity $R_{n,ab}$ is actually a dual superconformal invariant. We have already verified that it is invariant under dual conformal inversions. Since dual translation invariance and Lorentz invariance are manifest, the inversion symmetry is equivalent to invariance under the dual special conformal transformations, generated by $K_{\a \adt}$. 
It remains to show that it is invariant under the fermionic generators. Invariance under the chiral dual supersymmetry $Q_{\a A}$ is manifest (the $\theta$ variables only appear as differences in $R_{n,ab}$) and hence by commutation with $K_{\a \adt}$ we know that $\bar{S}_{\adt A}$ is a symmetry. The non-trivial symmetry to verify is the anti-chiral dual supersymmetry $\bar{Q}_\adt^A$.

To show that $\bar{Q}_{\adt}^{A}$ is indeed a symmetry of $R_{n,ab}$ we can exploit $Q$ and $\bar{S}$ by using a finite transformation made from these generators to fix a frame where $\theta_a=\theta_b=0$. Since $R_{n,ab}$ is invariant under $Q$ and $\bar{S}$ and all generators which arise through commutation of these with $\bar{Q}$ we know that $\bar{Q} R_{n,ab}$ is invariant under $Q$ and $\bar{S}$. So we can evaluate $\bar{Q} R_{n,ab}$ in the frame where $\theta_a=\theta_b=0$,
\begin{align}
\bar{Q}_{\adt}^{A} R_{n,ab} &= \theta_{n}^{\a A} \frac{\partial}
{\partial x_{n}^{\a \adt} } \biggl( \frac{ \l< a \, a-1\r> \l< b \, b-1\r> \delta^4\bigl( \l< n \theta_n \r> \bigr) }{ x_{ab}^2 \l< n | x_{na} x_{ab} | b \r> \l< n |x_{na} x_{ab} | b-1\r> \l<n|x_{nb}x_{ba} | a \r> \l< n |x_{nb} x_{ba} |a-1\r>}
 \biggr). \notag \\
 & \propto \l< n \theta_n \r> \delta^4\bigl( \l< n \theta_n \r>\bigr) = 0.
\end{align}
Thus we can see that the nilpotent nature of $R_{n,ab}$ is crucial in satisfying the invariance.

As we have seen the higher $R$-invariants appearing in the tree-level S-matrix are also dual conformal invariants. They are not dual superconformal invariants, as they are not annihilated by $\bar{Q}$. However they always appear in a nested fashion. For example at NNMHV level the quantity $R_{n,b_1 a_1, a_2 b_2}$ always appears multiplied by $R_{n,a_1 b_1}$. The $\bar{Q}$ variation of $R_{n;b_1 a_1,a_2 b_2}$ vanishes when multiplied by $R_{n,a_1b_1}$ so that the product is again dual superconformal invariant. Again we refer the reader to \cite{Drummond:2008cr} for more details.

The end result for the full superamplitude is that the function $\mathcal{P}_n$ is invariant under dual superconformal symmetry while the MHV prefactor is covariant under $D$, $C$, $K$ and $S$ and invariant under all other dual superconformal transformations. Thus we have
\be
D \mathcal{A}_n = n \mathcal{A}_n, \qquad C \mathcal{A}_n = n \mathcal{A}_n, \qquad  K^{\a \adt} \mathcal{A}_n = -\sum_i x_i^{\a \adt} \mathcal{A}_n, \qquad S^{\a A} \mathcal{A}_n = -\sum_i \theta_i^{\a A} \mathcal{A}_n\,.
\label{dualsconfcov}
\ee
We can get some insight into the nature of the symmetries we have found by combining the dual superconformal symmetry with the original one. It is particularly simple to describe the symmetry in terms of twistor variables. In $(2,2)$ signature the twistor variables are simply related to the on-shell superspace variables $(\lambda,\tilde{\lambda}|\eta)$ by a Fourier transformation $\lambda \longrightarrow \tilde{\mu}$ \cite{Witten:2003nn}. We then denote the (super)twistors as $\cZ_i^{\AA} = (Z_i^{A'} | \eta_i^A) = (\tilde{\mu}_i^{\alpha},\tilde{\lambda}_i^{\dot\alpha}|\eta_i^A)$.
Twistor space linearises the action of the superconformal generators $j^{\AA}{}_{\BB}$. The dual superconformal symmetry is then equivalent to introducing another set of `level-one' generators \cite{Drummond:2009fd},
\begin{align}
j^{\AA}{}_{\BB} &= \sum_i \cZ_i^{\AA} \frac{\partial}{\partial \cZ_i^{\BB}}, \label{twistorsconf}\\
j^{(1)}{}^{\AA}{}_{\BB} &= \sum_{i<j} (-1)^{\CC}\Bigl[\cZ_i^{\AA} \frac{\partial}{\partial \cZ_i^{\CC}} \cZ_j^{\CC} \frac{\partial}{\partial \cZ_j^{\BB}} - (i,j) \Bigr].
\label{twistoryangian}
\end{align}
Both of the formulas (\ref{twistorsconf}) and (\ref{twistoryangian}) are understood to have the supertrace 
removed. In this representation the generators of superconformal symmetry are first-order operators while the level-one Yangian generators are second order. The statement of the symmetry is then the following,
\be
j^{\AA}{}_{\BB} \mathcal{A}_n = 0\,,\qquad j^{(1)}{}^{\AA}{}_{\BB} \mathcal{A}_n =0\,.
\ee
This representation of the Yangian also makes an appearance in \cite{Dolan:2003uh,Dolan:2004ps} and provides an explanation for degeneracies in the spectrum of anomalous dimensions observed at the leading orders in perturbation theory.

In fact one can also phrase things the other way round. In \cite{Drummond:2010qh} it was demonstrated that there exists an alternative T-dual representation of the symmetry, where it is the dual superconformal generators $J_a$ which play the role of the level-zero generators, while the generators $k$ and $s$ of the original superconformal symmetry form part of the level-one generators. The generators take the same form as (\ref{twistorsconf}) and (\ref{twistoryangian}) but with the twistor variables replaced by momentum twistor variables \cite{Hodges:2009hk}. Momentum twistors are the twistors associated to the dual space and linearise the dual superconformal symmetry. They are defined as $\mathcal{W}_i^{\AA} = (W_i^{A'} | \chi_i^A)=(\lambda_i^\a,\mu_i^\adt | \chi_i^A)$ with
\be
\mu_i^\a = x^{\a \adt}_i \lam_{i\a}, \qquad \chi_i^A = \theta^{\a A}_i \lam_{i \a}.
\ee
The generators then take the form
\begin{align}
J^{\AA}{}_{\BB} &= \sum_i \cW_i^{\AA} \frac{\partial}{\partial \cW_i^{\BB}}, \label{momtwistordsconf}\\
J^{(1)}{}^{\AA}{}_{\BB} &= \sum_{i<j} (-1)^{\CC}\Bigl[\cW_i^{\AA} \frac{\partial}{\partial \cW_i^{\CC}} \cW_j^{\CC} \frac{\partial}{\partial \cW_j^{\BB}} - (i,j) \Bigr],
\label{momtwistoryangian}
\end{align}
again with supertraces removed. In this representation the generators annihilate the amplitude with the MHV prefactor dropped, i.e.
\be
J_a \mathcal{P}_n = 0, \qquad J^{(1)}_a \mathcal{P}_n = 0.
\ee

It is instructive to write the simplest R-invariant that we have seen in terms of momentum twistor variables. We refer to \cite{Mason:2009qx} for the details. We have
\be
R_{n,ab} = \frac{\delta^{0|4}(\chi_n (a-1\,\,a\, b-1\,\,b) + {\rm cyc}(n,a-1,a,b-1,b))}{(n\, a-1\,\,a\,b-1\,\,b)(a-1\,\,a\, b-1\,\,b) (a\, b-1\,\,b\, n)(b-1\,\,b\, n\, a-1)(b\, n\, a-1\,\,a)}\,.
\label{Rnabmomtwist}
\ee
Here we used four-brackets defined by
\be
 (abcd) = \epsilon_{A'B'C'D'}W_a^{A'}W_b^{B'}W_c^{C'}W_d^{D'}\,.
\ee
In the form (\ref{Rnabmomtwist}) dual superconformal symmetry is more transparent (that is a benefit of working with momentum twistors). It is also obviously a special case of a more general invariant,
\be
R(a,b,c,d,e) = \frac{\delta^{0|4}(\chi_a (bcde) + {\rm cyc}(a,b,c,d,e))}{(abcd)(bcde)(cdea)(deab)(eabc)}\,.
\ee
The existence of such quantities raises the question as to their role in the theory. In general invariants of the large symmetry algebra give `leading singularities' of the scattering amplitudes.



\section{Leading singularities and the loop integrand}

We have seen that the tree-level amplitude is a particular linear combination of Yangian invariants which is generated by the BCFW recursion relations. It is natural to ask what role the different individual invariants play in the theory and indeed if there is a role for the more general invariants appearing above. Indeed there is a natural class of objects which go by the name of leading singularities.

The best way to think of these objects is to consider the loop amplitude expanded in the coupling constant. At $l$ loops the amplitude will be the integral over $l$ internal loop momenta of some rational function $R_l$ of the external momenta and the loop momenta,
\be
\mathcal{A}_n = \sum_l a^l \int d^{4l}k R_l(p_1,\ldots,p_n;k_1,\ldots,k_l)\,.
\label{allloopamp}
\ee
The integrand $R$ in the above expression is only canonically defined in a planar theory. For general non-planar diagrams it is not clear how to combine naturally all contributions into a single overall integration. For planar diagrams however we can express the loop integrations as integrations over internal dual points and the function $R$ is taken to be completely symmetric under the exchange of the integration points, allowing all terms to be written as a single $l$-loop integral. These and other aspects of the loop integrand are discussed in detail in \cite{ArkaniHamed:2010kv}.

If we take the above expression and perform the the loop integrations over a compact contour instead of the usual non-compact real Minkowski space contour then we obtain a well-defined (and infrared finite) expression which is some algebraic function of the external momenta. The function is algebraic because the compact integrations can all be done by enclosing poles and using the residue theorem thus the internal momenta are all fixed by algebraic equations in terms of the external ones. The resulting function, $I_{n}(C)$ will depend on the choice of integration cycle and is called a leading singularity,
\be
I_{n,l}(C) = \int_C d^{4l}k R_l(p_1,\ldots,p_n;k_1,\ldots,k_l)\,.
\ee

\begin{figure}
\psfrag{a}[cc][cc]{\parbox[t]{0mm}{${}$}}
\psfrag{b}[cc][cc]{\parbox[t]{0mm}{${}$}}
\psfrag{c}[cc][cc]{\parbox[t]{0mm}{${}$}}
\psfrag{d}[cc][cc]{\parbox[t]{0mm}{${}$}}
\psfrag{e}[cc][cc]{\parbox[t]{0mm}{${}$}}
\psfrag{f}[cc][cc]{\parbox[t]{0mm}{${}$}}
\psfrag{g}[cc][cc]{\parbox[t]{0mm}{${}$}}
\psfrag{h}[cc][cc]{\parbox[t]{0mm}{${}$}}
 \centerline{{\epsfysize6cm
\epsfbox{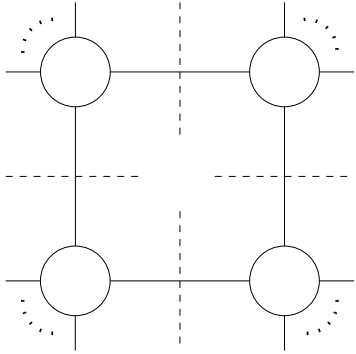}}}  \caption[]{\small The four-particle cuts of one-loop amplitudes correspond to a sum of two leading singularities, given by the two solutions of the on-shell constraints.}
  \label{4mass}
\end{figure}

Leading singularities can also be thought of as being constructed from tree-level data. A simple example of this idea comes from the generalised four-particle cuts of \cite{Britto:2004nc}. Here we consider one-loop amplitudes with the loop integration performed over a contour that localises it so that four internal propagators in the loop go on shell. The amplitude is then separated into four quarters, as illustrated in Fig. \ref{4mass}, with each quarter described an on-shell tree-level amplitude. As for the BCFW diagrams, there is a sum over exchanged states on the internal lines, captured by a Grassmann integration in the $\mathcal{N}=4$ theory \cite{Drummond:2008bq,Brandhuber:2008pf,ArkaniHamed:2008gz}. Note that there are two possible leading singularities associated with a given four-particle cut, corresponding to the two solutions of the cut conditions. The leading singularities are chiral in that one solution may contribute to the MHV amplitude while the other contributes to $\overline{\rm MHV}$, for example.

From the point of view of the symmetry of the theory, it is possible to see that the leading singularities are also invariants. To see this one can generalise the arguments of \cite{Drummond:2008bq,Brandhuber:2008pf} to higher loops as in \cite{Drummond:2010mb} or, very elegantly, phrase the gluing operations of tree-level amplitudes in terms of Yangian-invariant operations \cite{ArkaniHamed:2010kv}. The Yangian invariance of all leading singularities is equivalent to the statement that the loop integrand is itself invariant up to a total derivative,
\be
\delta I_{n,l}(C) =0 \quad \forall\, C \iff \delta d^{4l}k R_l(p_1,\ldots,p_n;k_1,\ldots,k_l) = d X\,.
\ee

In \cite{ArkaniHamed:2009dn} a remarkable formula was proposed which computes leading singularities of scattering amplitudes in $\mathcal{N}=4$ super Yang-Mills theory. The formula takes the form of an integral over the Grassmannian $G(k,n)$, the space of complex $k$-planes in $\mathbb{C}^n$. The integrand is a specific $k(n-k)$-form $K$ to be integrated over cycles $C$ of the corresponding dimension, with the integral being treated as a multi-dimensional contour integral.
The result obtained depends on the choice of contour and is non-vanishing for closed contours because the form has poles located on certain hyperplanes in the Grassmannian,
\begin{equation}
\mathcal{L} = \int_C K.
\end{equation}

The form $K$ is constructed from a product of superconformally-invariant delta functions
of linear combinations of twistor variables. It is through this factor that the integral depends on the kinematic data of the $n$-point scattering amplitude of the gauge theory. The delta functions are multiplied by a cyclically invariant function on the Grassmannian which has poles. Specifically the formula takes the following form in twistor space
\begin{equation}
\mathcal{L}_{\rm ACCK}(\cZ) = \int \frac{D^{k(n-k)}c }{\mathcal{M}_1 \ldots \mathcal{M}_n} \prod_{a=1}^k \delta^{4|4}\Bigl(\sum_{i=1}^n c_{ai} \cZ_i\Bigr) ,
\label{ACCK}
\end{equation}
where the $c_{ai}$ are complex parameters which are integrated choosing a specific contour. The form $D^{k(n-k)}c$ is the natural holomorphic globally $gl(n)$-invariant and locally $sl(k)$-invariant $(k(n-k),0)$-form given explicitly in \cite{Mason:2009qx}.
The denominator is the cyclic product of consecutive $(k \times k)$ minors $\mathcal{M}_p$  made from the columns $
p,\ldots,p+k-1$ of the $(k \times n)$ matrix of the $c_{ai}$
\begin{equation}
\mathcal{M}_{p} \equiv (p~p+1~p+2\ldots p+k-1)  .
\end{equation}
As described in \cite{ArkaniHamed:2009dn} the formula (\ref{ACCK}) has a $GL(k)$ gauge symmetry which implies that $k^2$ of the $c_{ai}$ are gauge degrees of freedom and therefore should not be integrated over. The remaining $k(n-k)$ are the true coordinates on the Grassmannian. 
This formula (\ref{ACCK}) produces leading singularities of N${}^{k-2}$MHV scattering amplitudes when suitable closed integration contours are chosen. This fact was explicitly verified up to eight points in \cite{ArkaniHamed:2009dn} and it was conjectured that the formula produces all possible leading singularities at all orders in the perturbative expansion.

The formula (\ref{ACCK}) has a T-dual version \cite{Mason:2009qx}, expressed in terms of momentum twistors. The momentum twistor Grassmannian formula takes the same form as the original
\begin{equation}
\mathcal{L}_{\rm MS}(\mathcal{W}) = \int \frac{D^{k(n-k)}t}{\mathcal{M}_1 \ldots \mathcal{M}_n} \prod_{a=1}^k  \delta^{4|4}\Bigl(\sum_{i=1}^n t_{ai} \cW_i\Bigr) ,
\label{MS}
\end{equation}
but now it is the dual superconformal symmetry that is manifest. The integration variables $t_{ai}$ are again a $(k \times n)$ matrix of complex parameters and we use the notation $\mathcal{M}_p$ to refer to $(k \times k)$ minors made from the matrix of the $t_{ai}$. The formula (\ref{MS}) produces the same objects as (\ref{ACCK}) but now with the MHV tree-level amplitude factored out. They therefore contribute to N${}^k$MHV amplitudes.

The equivalence of the two formulations (\ref{ACCK}) and (\ref{MS}) was shown in \cite{ArkaniHamed:2009vw} via a change of variables. Therefore, since each of the formulas has a different superconformal symmetry manifest, they both possess an invariance under the Yangian $Y(psl(4|4))$. The Yangian symmetry of these formulas was explicitly demonstrated in \cite{Drummond:2010qh} by directly applying the Yangian level-one generators to the Grassmannian integral itself.


In \cite{Drummond:2010qh} it was found that applying the level-one generator to the form $K$ yields a total derivative,
\begin{equation}
J^{(1)}{}^{\AA}{}_{\BB} K = d \Omega^{\AA}{}_{\BB}.
\label{exactvar}
\end{equation}
This property guarantees that $\mathcal{L}$ is invariant for every choice of closed contour. Moreover it has been shown \cite{Drummond:2010uq,Korchemsky:2010ut} that the form $K$ is unique after imposing the condition (\ref{exactvar}). In this sense the Grassmannian integral is the most general form of Yangian invariant. Moreover, replacing $\delta^{4|4} \longrightarrow \delta^{m|m}$, the formulas (\ref{ACCK},\ref{MS}) are equally valid for generating invariants of the symmetry $Y(psl(m|m))$ where one no longer has the interpretation of the symmetry as superconformal symmetry. Thus the Grassmannian integral formula is really naturally associated to the series of Yangians $Y(psl(m|m))$.


Finally, let us note that the recursive arguments we saw at tree-level can also be applied to the loop integrand \cite{ArkaniHamed:2010kv,Boels:2010nw}. In \cite{ArkaniHamed:2010kv} a specific recursion relation was proposed for the all-loop integrand of planar $\mathcal{N}=4$ super Yang-Mills theory. It reads,
\begin{align}
M_{n,l}(1,\ldots,n) =& M_{n-1,l}(1,\ldots,n-1) + \sum_{n_L,j} M_{n_R,l_R}(1,\ldots,j,I_j) M_{n_L,l_L}(I_j,j+1,\ldots,\hat{n}_j)\notag \\
 +& \int_{\rm GL(2)} R(A,B,n-1,n,1) M_{n+2,l-1}(1,\ldots,\hat{n}_{AB},\hat{A},B)\,.
\end{align}
Here the $n$-particle $l$-loop (super)integrand is denoted $M_{n,l}$. As well as the explicit dependence on the external kinematical variables (here taken to be the momentum twistors) it depends on $l$ internal loop integration points (or lines $(AB)_k$ in momentum twistor space). The summation is performed so that $n_L+n_R=n+2$ and $l_L+l_R=l$, while the various momentum twistors entering the formula are defined by,
\begin{align}
&\hat{n}_j = (n-1\,n)\cap (j\,j+1\,1), \qquad I_j = (j\, j+1) \cap (n-1\, n\, 1), \notag\\
&\hat{n}_{AB} = (n-1 \, n)\cap (AB1), \qquad \,\,\, \hat{A}=(AB)\cap (n-1\, n\, 1)\,.
\end{align}
The recursion relation is very similar in form to (\ref{fullPrec}) but there is an additional term involving a $GL(2)$ integration. This term corresponds to taking forward limit of an $l-1$ loop integrand - it essentially corresponds to the possibility of one of the loop propagators going on shell as a result of the shift. The integration is performed over the $GL(2)$ in order to result in an expression with a $GL(2)$ invariance in the variables $A$ and $B$ so that the final term only depends on the line $(AB)$. Finally one should symmetrise over all internal integration variables (or momentum twistor lines). This recursive relation can also be phrased in terms of Yangian-invariant operations \cite{ArkaniHamed:2010kv}, showing that the integrand is invariant up to total derivative. The integrand obtained from this recursion relation is four-dimensional in nature and unregulated. Moreover, just as in the case of the tree-level recursion relation, the singularities of each term are of a non-local nature. In order to perform the loop integrations canonically the expressions should first be reorganised so that the non-local poles cancel manifestly, leaving only the local propagators \cite{ArkaniHamed:2010kv,ArkaniHamed:2010gh}. Then, following \cite{Alday:2009zm}, one can regularise the infrared divergences by introducing a mass. See \cite{Henn:2010bk,Henn:2010ir,Drummond:2010mb} for explicit calculations of such amplitudes and \cite{Henn:2011xk} for a discussion of these ideas in this review series.

\section{Dual conformal symmetry in higher dimensions}

Some of the symmetry we have seen is in fact not confined to the four-dimensional incarnation of maximally supersymmetric gauge theories. One can consider also higher-dimensional versions, the $\mathcal{N}=(1,1)$ super Yang-Mills theory in six dimensions, or the $\mathcal{N}=1$ theory in ten dimensions. In both cases the tree-level amplitudes exhibit dual conformal symmetry \cite{Bern:2010qa,Dennen:2010dh,CaronHuot:2010rj}. For example, in six dimensions, following the spinor-helicity conventions of \cite{Cheung:2009dc}, we can write the particle momenta as,
\be
p_i^{AB} = \lambda_i^{Aa} \epsilon_{ab} \lambda_i^{Bb}\,, \qquad p_{iAB} = \bar{\lambda}_{iA\dot{a}} \epsilon^{\dot{a}\dot{b}} \bar{\lambda}_{iB\dot{b}}\,, \qquad p_{iAB} = \tfrac{1}{2} p_i^{CD} \epsilon_{ABCD}\,.
\ee
Here the capital indices are fundamental/anti-fundamental $Spin(1,5)$ indices, running from $1$ to $4$ while the indices $a$ and $\dot{a}$ are the $SU(2)\times SU(2)$ little group labels. 
The on-shell $\mathcal{N}=(1,1)$ supermultiplet can be organised into the following on-shell superfield,
\be
\Phi(\eta,\tilde{\eta}) = \phi + \chi^a \eta_a  + \tilde{\chi}_{\dot{a}} \tilde{\eta}^{\dot{a}} + \phi' (\eta)^2 + g^a_{\dot a} \eta_a \tilde{\eta}^{\dot{a}} + \phi'' (\tilde{\eta})^2 + \tilde{\psi}_{\dot{a}} \tilde{\eta}^{\dot{a}} (\eta)^2 + \psi^a \eta_a (\tilde{\eta})^2 + \phi''' (\eta)^2 (\tilde{\eta})^2\,.
\ee
The multiplicative supercharges take the form
\be
q_i^{A} = \lambda_i^{Aa}\eta_{ia}\,, \qquad \tilde{q}_{iA} = \bar{\lambda}_{iA\dot{a}} \tilde{\eta}_i^{\dot{a}}\,.
\ee
Now one can introduce six-dimensional dual coordinates following the four-dimensional case,
\be
x_i^{AB} - x_{i+1}^{AB} = p_i^{AB}, \qquad \theta_i^A -\theta_{i+1}^A = \lambda_i^{Aa} \eta_{ia}, \qquad \bar{\theta}_{iA} - \bar{\theta}_{i+1A} = \bar{\lambda}_{iA\dot{a}}\tilde{\eta}_i^{\dot{a}}\,.
\label{dualcoords6d}
\ee
These variables transform canonically under dual conformal inversions, 
\be
x_i^\mu \longrightarrow \frac{x_i^\mu}{x_i^2},\qquad \theta_i^A \longrightarrow (x_i^{-1})_{AB}\theta_i^B\,, \qquad \bar{\theta}_{iA} \longrightarrow (x_i^{-1})^{AB} \bar{\theta}_{iB}\,.
\ee
and the relations (\ref{dualcoords6d}) define corresponding transformations of the $\lambda_i$ etc. The tree level amplitudes are again covariant, when suitably stripped of the momentum and supersymmetry delta functions. We define
\be
\mathcal{A}_n  = \delta^6(p) \delta^4(q) \delta^4(\tilde{q}) f_n
\ee
and the dual conformal transformation of $f_n$ is simple,
\be
f_n \longrightarrow (x_1^2 \ldots x_n^2) f_n\,.
\ee
This statement is a non-trivial constraint on the form of the amplitudes in six-dimensional super Yang-Mills theory.
It can again be seen recursively from the BCFW recursion relations for the tree-level amplitudes. Note that in dimensions higher than four, Yang-Mills theory is not a conformal field theory and so we do not expect the analogue of the original conformal symmetry here. 
The six-dimensional amplitudes can be related \cite{Bern:2010qa,Dennen:2010dh} to four-dimensional ones on the Coulomb branch as studied in \cite{Alday:2009zm} and the presence of the dual conformal symmetry in these two cases is related. The induced masses of the external momenta on the Coulomb branch are related to the fifth and sixth components of the massless six-dimensional momenta.  

In ten dimensions, dual conformal symmetry again appears for the tree-level amplitudes. In that case an analogue of the level-one momentum generator is defined in order to show the invariance. We refer the reader to \cite{CaronHuot:2010rj} for more details.
\section*{Acknowledgements}

It is a pleasure to thank Livia Ferro, Johannes Henn, Gregory Korchemsky, Jan Plefka and Emery Sokatchev for collaboration on the subjects covered in this review article.

\bibliography{ampsreview}

\begin{thebibliography}{10}
\ifx\href\asklfhas\newcommand{\href}[2]{#2}\fi
\ifx\arxivref\asklfhas\newcommand{\arxivref}[2]{\href{http://arxiv.org/abs/#1}{#2}}\fi
\ifx\doiref\asklfhas\newcommand{\doiref}[2]{\href{http://dx.doi.org/#1}{#2}}\fi
\raggedright
\small
\parskip 0pt

\bibitem{Bern:1994zx}
Z.~Bern, L.~J.~Dixon, D.~C.~Dunbar and D.~A.~Kosower,
\textit{``{One-Loop n-Point Gauge Theory Amplitudes, Unitarity and Collinear
  Limits}''},
\textsf{\doiref{10.1016/0550-3213(94)90179-1}{Nucl.~Phys.~B425,~217~(1994)}},
\texttt{\arxivref{hep-ph/9403226}{hep-ph/9403226}}.

\bibitem{Bern:1994cg}
Z.~Bern, L.~J.~Dixon, D.~C.~Dunbar and D.~A.~Kosower,
\textit{``{Fusing gauge theory tree amplitudes into loop amplitudes}''},
\textsf{\doiref{10.1016/0550-3213(94)00488-Z}{Nucl.~Phys.~B435,~59~(1995)}},
\texttt{\arxivref{hep-ph/9409265}{hep-ph/9409265}}.

\bibitem{Britto:2004nc}
R.~Britto, F.~Cachazo and B.~Feng,
\textit{``{Generalized unitarity and one-loop amplitudes in $\mathcal{N}$ = 4
  super-Yang-Mills}''},
\textsf{\doiref{10.1016/j.nuclphysb.2005.07.014}{Nucl.~Phys.~B725,~275~(2005)}},
\texttt{\arxivref{hep-th/0412103}{hep-th/0412103}}.

\bibitem{Bern:2011qt}
Z.~Bern and Y.-t.~Huang,
\textit{``{Basics of Generalized Unitarity}''},
\texttt{\arxivref{1103.1869}{arxiv:1103.1869}}.

\bibitem{Britto:2004ap}
R.~Britto, F.~Cachazo and B.~Feng,
\textit{``{New Recursion Relations for Tree Amplitudes of Gluons}''},
\textsf{\doiref{10.1016/j.nuclphysb.2005.02.030}{Nucl.~Phys.~B715,~499~(2005)}},
\texttt{\arxivref{hep-th/0412308}{hep-th/0412308}}.

\bibitem{Britto:2005fq}
R.~Britto, F.~Cachazo, B.~Feng and E.~Witten,
\textit{``{Direct Proof Of Tree-Level Recursion Relation In Yang- Mills
  Theory}''},
\textsf{\doiref{10.1103/PhysRevLett.94.181602}{Phys.~Rev.~Lett.~94,~181602~(2005)}},
\texttt{\arxivref{hep-th/0501052}{hep-th/0501052}}.

\bibitem{Witten:2003nn}
E.~Witten,
\textit{``{Perturbative gauge theory as a string theory in twistor space}''},
\textsf{\doiref{10.1007/s00220-004-1187-3}{Commun.~Math.~Phys.~252,~189~(2004)}},
\texttt{\arxivref{hep-th/0312171}{hep-th/0312171}}.

\bibitem{Drummond:2008vq}
J.~M.~Drummond, J.~Henn, G.~P.~Korchemsky and E.~Sokatchev,
\textit{``{Dual superconformal symmetry of scattering amplitudes in
  $\mathcal{N}$ = 4 super-Yang-Mills theory}''},
\texttt{\arxivref{0807.1095}{arxiv:0807.1095}}.

\bibitem{Drummond:2009fd}
J.~M.~Drummond, J.~M.~Henn and J.~Plefka,
\textit{``{Yangian symmetry of scattering amplitudes in $\mathcal{N}$ = 4 super
  Yang-Mills theory}''},
\textsf{\doiref{10.1088/1126-6708/2009/05/046}{JHEP~0905,~046~(2009)}},
\texttt{\arxivref{0902.2987}{arxiv:0902.2987}}.

\bibitem{Berkovits:2008ic}
N.~Berkovits and J.~Maldacena,
\textit{``{Fermionic T-Duality, Dual Superconformal Symmetry, and the
  Amplitude/Wilson Loop Connection}''},
\textsf{\doiref{10.1088/1126-6708/2008/09/062}{JHEP~0809,~062~(2008)}},
\texttt{\arxivref{0807.3196}{arxiv:0807.3196}}.

\bibitem{Beisert:2008iq}
N.~Beisert, R.~Ricci, A.~A.~Tseytlin and M.~Wolf,
\textit{``{Dual Superconformal Symmetry from AdS$_5$ $\times$ S$^5$ Superstring
  Integrability}''},
\textsf{\doiref{10.1103/PhysRevD.78.126004}{Phys.~Rev.~D78,~126004~(2008)}},
\texttt{\arxivref{0807.3228}{arxiv:0807.3228}}.

\bibitem{Beisert:2010jr}
N.~Beisert et~al.,
\textit{``{Review of AdS/CFT Integrability: An Overview}''},
\texttt{\arxivref{1012.3982}{arxiv:1012.3982}}.

\bibitem{ArkaniHamed:2010kv}
N.~Arkani-Hamed, J.~L.~Bourjaily, F.~Cachazo, S.~Caron-Huot and J.~Trnka,
\textit{``{The All-Loop Integrand For Scattering Amplitudes in Planar N=4
  SYM}''},
\texttt{\arxivref{1008.2958}{arxiv:1008.2958}}.

\bibitem{Beisert:2004ry}
N.~Beisert,
\textit{``{The dilatation operator of $\mathcal{N}$ = 4 super Yang-Mills theory
  and integrability}''},
\textsf{\doiref{10.1016/j.physrep.2004.09.007}{Phys.~Rept.~405,~1~(2005)}},
\texttt{\arxivref{hep-th/0407277}{hep-th/0407277}}.

\bibitem{Drummond:2008cr}
J.~M.~Drummond and J.~M.~Henn,
\textit{``{All tree-level amplitudes in $\mathcal{N}$ = 4 SYM}''},
\textsf{\doiref{10.1088/1126-6708/2009/04/018}{JHEP~0904,~018~(2009)}},
\texttt{\arxivref{0808.2475}{arxiv:0808.2475}}.

\bibitem{Brandhuber:2008pf}
A.~Brandhuber, P.~Heslop and G.~Travaglini,
\textit{``{A note on dual superconformal symmetry of the $\mathcal{N}$ = 4
  super Yang-Mills S-matrix}''},
\textsf{\doiref{10.1103/PhysRevD.78.125005}{Phys.~Rev.~D78,~125005~(2008)}},
\texttt{\arxivref{0807.4097}{arxiv:0807.4097}}.

\bibitem{ArkaniHamed:2008gz}
N.~Arkani-Hamed, F.~Cachazo and J.~Kaplan,
\textit{``{What is the Simplest Quantum Field Theory?}''},
\texttt{\arxivref{0808.1446}{arxiv:0808.1446}}.

\bibitem{Elvang:2008na}
H.~Elvang, D.~Z.~Freedman and M.~Kiermaier,
\textit{``{Recursion Relations, Generating Functions, and Unitarity Sums in N=4
  SYM Theory}''},
\textsf{\doiref{10.1088/1126-6708/2009/04/009}{JHEP~0904,~009~(2009)}},
\texttt{\arxivref{0808.1720}{arxiv:0808.1720}}.

\bibitem{Bargheer:2009qu}
T.~Bargheer, N.~Beisert, W.~Galleas, F.~Loebbert and T.~McLoughlin,
\textit{``{Exacting $\mathcal{N}$ = 4 Superconformal Symmetry}''},
\texttt{\arxivref{0905.3738}{arxiv:0905.3738}}.

\bibitem{Korchemsky:2009hm}
G.~P.~Korchemsky and E.~Sokatchev,
\textit{``{Symmetries and analytic properties of scattering amplitudes in
  $\mathcal{N}$ = 4 SYM theory}''},
\texttt{\arxivref{0906.1737}{arxiv:0906.1737}}.

\bibitem{Sever:2009aa}
A.~Sever and P.~Vieira,
\textit{``{Symmetries of the $\mathcal{N}$ = 4 SYM S-matrix}''},
\texttt{\arxivref{0908.2437}{arxiv:0908.2437}}.

\bibitem{Bargheer:2011mm}
T.~Bargheer, N.~Beisert and F.~Loebbert,
\textit{``{Exact Superconformal and Yangian Symmetry of Scattering
  Amplitudes}''},
\texttt{\arxivref{1104.0700}{arxiv:1104.0700}}.

\bibitem{Dixon:1996wi}
L.~J.~Dixon,
\textit{``{Calculating scattering amplitudes efficiently}''},
\texttt{\arxivref{hep-ph/9601359}{hep-ph/9601359}}.

\bibitem{Dixon:2011xs}
L.~J.~Dixon,
\textit{``{Scattering amplitudes: the most perfect microscopic structures in
  the universe}''},
\texttt{\arxivref{1105.0771}{arxiv:1105.0771}}.

\bibitem{Drummond:2010ep}
J.~M.~Drummond,
\textit{``{Hidden Simplicity of Gauge Theory Amplitudes}''},
\textsf{\doiref{10.1088/0264-9381/27/21/214001}{Class.~Quant.~Grav.~27,~214001~(2010)}},
\texttt{\arxivref{1010.2418}{arxiv:1010.2418}}.

\bibitem{Brandhuber:2011ke}
A.~Brandhuber, B.~Spence and G.~Travaglini,
\textit{``{Tree-Level Formalism}''},
\texttt{\arxivref{1103.3477}{arxiv:1103.3477}}.

\bibitem{ArkaniHamed:2008yf}
N.~Arkani-Hamed and J.~Kaplan,
\textit{``{On Tree Amplitudes in Gauge Theory and Gravity}''},
\textsf{\doiref{10.1088/1126-6708/2008/04/076}{JHEP~0804,~076~(2008)}},
\texttt{\arxivref{0801.2385}{arxiv:0801.2385}}.

\bibitem{Nair:1988bq}
V.~P.~Nair,
\textit{``{A current algebra for some gauge theory amplitudes}''},
\textsf{\doiref{10.1016/0370-2693(88)91471-2}{Phys.~Lett.~B214,~215~(1988)}}.

\bibitem{Bern:2004bt}
Z.~Bern, L.~J.~Dixon and D.~A.~Kosower,
\textit{``{All next-to-maximally helicity-violating one-loop gluon amplitudes
  in $\mathcal{N}$ = 4 super-Yang-Mills theory}''},
\textsf{\doiref{10.1103/PhysRevD.72.045014}{Phys.~Rev.~D72,~045014~(2005)}},
\texttt{\arxivref{hep-th/0412210}{hep-th/0412210}}.

\bibitem{Britto:2005dg}
R.~Britto, B.~Feng, R.~Roiban, M.~Spradlin and A.~Volovich,
\textit{``{All split helicity tree-level gluon amplitudes}''},
\textsf{\doiref{10.1103/PhysRevD.71.105017}{Phys.~Rev.~D71,~105017~(2005)}},
\texttt{\arxivref{hep-th/0503198}{hep-th/0503198}}.

\bibitem{Dixon:2010ik}
L.~J.~Dixon, J.~M.~Henn, J.~Plefka and T.~Schuster,
\textit{``{All tree-level amplitudes in massless QCD}''},
\textsf{\doiref{10.1007/JHEP01(2011)035}{JHEP~1101,~035~(2011)}},
\texttt{\arxivref{1010.3991}{arxiv:1010.3991}}.

\bibitem{Bedford:2005yy}
J.~Bedford, A.~Brandhuber, B.~J.~Spence and G.~Travaglini,
\textit{``{A recursion relation for gravity amplitudes}''},
\textsf{\doiref{10.1016/j.nuclphysb.2005.016}{Nucl.~Phys.~B721,~98~(2005)}},
\texttt{\arxivref{hep-th/0502146}{hep-th/0502146}}.

\bibitem{Cachazo:2005ca}
F.~Cachazo and P.~Svrcek,
\textit{``{Tree level recursion relations in general relativity}''},
\texttt{\arxivref{hep-th/0502160}{hep-th/0502160}}.

\bibitem{Drummond:2009ge}
J.~M.~Drummond, M.~Spradlin, A.~Volovich and C.~Wen,
\textit{``{Tree-Level Amplitudes in N=8 Supergravity}''},
\textsf{\doiref{10.1103/PhysRevD.79.105018}{Phys.~Rev.~D79,~105018~(2009)}},
\texttt{\arxivref{0901.2363}{arxiv:0901.2363}}.

\bibitem{Alday:2007hr}
L.~F.~Alday and J.~M.~Maldacena,
\textit{``{Gluon scattering amplitudes at strong coupling}''},
\textsf{\doiref{10.1088/1126-6708/2007/06/064}{JHEP~0706,~064~(2007)}},
\texttt{\arxivref{0705.0303}{arxiv:0705.0303}}.

\bibitem{Drummond:2007aua}
J.~M.~Drummond, G.~P.~Korchemsky and E.~Sokatchev,
\textit{``{Conformal properties of four-gluon planar amplitudes and Wilson
  loops}''},
\textsf{\doiref{10.1016/j.nuclphysb.2007.11.041}{Nucl.~Phys.~B795,~385~(2008)}},
\texttt{\arxivref{0707.0243}{arxiv:0707.0243}}.

\bibitem{Brandhuber:2007yx}
A.~Brandhuber, P.~Heslop and G.~Travaglini,
\textit{``{MHV Amplitudes in $\mathcal{N}$ = 4 Super Yang-Mills and Wilson
  Loops}''},
\textsf{\doiref{10.1016/j.nuclphysb.2007.11.002}{Nucl.~Phys.~B794,~231~(2008)}},
\texttt{\arxivref{0707.1153}{arxiv:0707.1153}}.

\bibitem{Drummond:2007cf}
J.~M.~Drummond, J.~Henn, G.~P.~Korchemsky and E.~Sokatchev,
\textit{``{On planar gluon amplitudes/Wilson loops duality}''},
\textsf{\doiref{10.1016/j.nuclphysb.2007.11.007}{Nucl.~Phys.~B795,~52~(2008)}},
\texttt{\arxivref{0709.2368}{arxiv:0709.2368}}.

\bibitem{Drummond:2007au}
J.~M.~Drummond, J.~Henn, G.~P.~Korchemsky and E.~Sokatchev,
\textit{``{Conformal Ward identities for Wilson loops and a test of the duality
  with gluon amplitudes}''},
\texttt{\arxivref{0712.1223}{arxiv:0712.1223}}.

\bibitem{Drummond:2007bm}
J.~M.~Drummond, J.~Henn, G.~P.~Korchemsky and E.~Sokatchev,
\textit{``{The hexagon Wilson loop and the BDS ansatz for the six-gluon
  amplitude}''},
\textsf{\doiref{10.1016/j.physletb.2008.03.032}{Phys.~Lett.~B662,~456~(2008)}},
\texttt{\arxivref{0712.4138}{arxiv:0712.4138}}.

\bibitem{Bern:2008ap}
Z.~Bern, L.~Dixon, D.~Kosower, R.~Roiban, M.~Spradlin, C.~Vergu and
  A.~Volovich,
\textit{``{The Two-Loop Six-Gluon MHV Amplitude in Maximally Supersymmetric
  Yang-Mills Theory}''},
\textsf{\doiref{10.1103/PhysRevD.78.045007}{Phys.~Rev.~D78,~045007~(2008)}},
\texttt{\arxivref{0803.1465}{arxiv:0803.1465}}.

\bibitem{Drummond:2008aq}
J.~M.~Drummond, J.~Henn, G.~P.~Korchemsky and E.~Sokatchev,
\textit{``{Hexagon Wilson loop = six-gluon MHV amplitude}''},
\textsf{\doiref{10.1016/j.nuclphysb.2009.02.015}{Nucl.~Phys.~B815,~142~(2009)}},
\texttt{\arxivref{0803.1466}{arxiv:0803.1466}}.

\bibitem{Anastasiou:2009kna}
C.~Anastasiou, A.~Brandhuber, P.~Heslop, V.~V.~Khoze, B.~Spence and
  G.~Travaglini,
\textit{``{Two-Loop Polygon Wilson Loops in $\mathcal{N}$ = 4 SYM}''},
\textsf{\doiref{10.1088/1126-6708/2009/05/115}{JHEP~0905,~115~(2009)}},
\texttt{\arxivref{0902.2245}{arxiv:0902.2245}}.

\bibitem{Drummond:2010km}
J.~M.~Drummond,
\textit{``{Review of AdS/CFT Integrability, Chapter V.2: Dual Superconformal
  Symmetry}''},
\texttt{\arxivref{1012.4002}{arxiv:1012.4002}}.

\bibitem{Ricci:2007eq}
R.~Ricci, A.~A.~Tseytlin and M.~Wolf,
\textit{``{On T-Duality and Integrability for Strings on AdS Backgrounds}''},
\textsf{\doiref{10.1088/1126-6708/2007/12/082}{JHEP~0712,~082~(2007)}},
\texttt{\arxivref{0711.0707}{arxiv:0711.0707}}.

\bibitem{Dolan:2003uh}
L.~Dolan, C.~R.~Nappi and E.~Witten,
\textit{``{A relation between approaches to integrability in superconformal
  Yang-Mills theory}''},
\textsf{\doiref{10.1088/1126-6708/2003/10/017}{JHEP~0310,~017~(2003)}},
\texttt{\arxivref{hep-th/0308089}{hep-th/0308089}}.

\bibitem{Dolan:2004ps}
L.~Dolan, C.~R.~Nappi and E.~Witten,
\textit{``{Yangian symmetry in D=4 superconformal Yang-Mills theory}''},
\texttt{\arxivref{hep-th/0401243}{hep-th/0401243}}.

\bibitem{Drummond:2010qh}
J.~M.~Drummond and L.~Ferro,
\textit{``{Yangians, Grassmannians and T-duality}''},
\texttt{\arxivref{1001.3348}{arxiv:1001.3348}}.

\bibitem{Hodges:2009hk}
A.~Hodges,
\textit{``{Eliminating spurious poles from gauge-theoretic amplitudes}''},
\texttt{\arxivref{0905.1473}{arxiv:0905.1473}}.

\bibitem{Mason:2009qx}
L.~Mason and D.~Skinner,
\textit{``{Dual Superconformal Invariance, Momentum Twistors and
  Grassmannians}''},
\texttt{\arxivref{0909.0250}{arxiv:0909.0250}}.

\bibitem{Drummond:2008bq}
J.~M.~Drummond, J.~Henn, G.~P.~Korchemsky and E.~Sokatchev,
\textit{``{Generalized unitarity for $\mathcal{N}$ = 4 super-amplitudes}''},
\texttt{\arxivref{0808.0491}{arxiv:0808.0491}}.

\bibitem{Drummond:2010mb}
J.~M.~Drummond and J.~M.~Henn,
\textit{``{Simple loop integrals and amplitudes in N=4 SYM}''},
\texttt{\arxivref{1008.2965}{arxiv:1008.2965}}.

\bibitem{ArkaniHamed:2009dn}
N.~Arkani-Hamed, F.~Cachazo, C.~Cheung and J.~Kaplan,
\textit{``{A Duality For The S Matrix}''},
\texttt{\arxivref{0907.5418}{arxiv:0907.5418}}.

\bibitem{ArkaniHamed:2009vw}
N.~Arkani-Hamed, F.~Cachazo and C.~Cheung,
\textit{``{The Grassmannian Origin Of Dual Superconformal Invariance}''},
\textsf{\doiref{10.1007/JHEP03(2010)036}{JHEP~1003,~036~(2010)}},
\texttt{\arxivref{0909.0483}{arxiv:0909.0483}}.

\bibitem{Drummond:2010uq}
J.~M.~Drummond and L.~Ferro,
\textit{``{The Yangian origin of the Grassmannian integral}''},
\texttt{\arxivref{1002.4622}{arxiv:1002.4622}}.

\bibitem{Korchemsky:2010ut}
G.~P.~Korchemsky and E.~Sokatchev,
\textit{``{Superconformal invariants for scattering amplitudes in N=4 SYM
  theory}''},
\texttt{\arxivref{1002.4625}{arxiv:1002.4625}}.

\bibitem{Boels:2010nw}
R.~H.~Boels,
\textit{``{On BCFW shifts of integrands and integrals}''},
\textsf{\doiref{10.1007/JHEP11(2010)113}{JHEP~1011,~113~(2010)}},
\texttt{\arxivref{1008.3101}{arxiv:1008.3101}}.

\bibitem{ArkaniHamed:2010gh}
N.~Arkani-Hamed, J.~L.~Bourjaily, F.~Cachazo and J.~Trnka,
\textit{``{Local Integrals for Planar Scattering Amplitudes}''},
\texttt{\arxivref{1012.6032}{arxiv:1012.6032}}.

\bibitem{Alday:2009zm}
L.~F.~Alday, J.~M.~Henn, J.~Plefka and T.~Schuster,
\textit{``{Scattering into the fifth dimension of $\mathcal{N}$ = 4 super
  Yang-Mills}''},
\texttt{\arxivref{0908.0684}{arxiv:0908.0684}}.

\bibitem{Henn:2010bk}
J.~M.~Henn, S.~G.~Naculich, H.~J.~Schnitzer and M.~Spradlin,
\textit{``{Higgs-regularized three-loop four-gluon amplitude in N=4 SYM:
  exponentiation and Regge limits}''},
\textsf{\doiref{10.1007/JHEP04(2010)038}{JHEP~1004,~038~(2010)}},
\texttt{\arxivref{1001.1358}{arxiv:1001.1358}}.

\bibitem{Henn:2010ir}
J.~M.~Henn, S.~G.~Naculich, H.~J.~Schnitzer and M.~Spradlin,
\textit{``{More loops and legs in Higgs-regulated N=4 SYM amplitudes}''},
\textsf{\doiref{10.1007/JHEP08(2010)002}{JHEP~1008,~002~(2010)}},
\texttt{\arxivref{1004.5381}{arxiv:1004.5381}}.

\bibitem{Henn:2011xk}
J.~M.~Henn,
\textit{``{Dual conformal symmetry at loop level: massive regularization}''},
\texttt{\arxivref{1103.1016}{arxiv:1103.1016}}.

\bibitem{Bern:2010qa}
Z.~Bern, J.~J.~Carrasco, T.~Dennen, Y.-t.~Huang and H.~Ita,
\textit{``{Generalized Unitarity and Six-Dimensional Helicity}''},
\texttt{\arxivref{1010.0494}{arxiv:1010.0494}}.

\bibitem{Dennen:2010dh}
T.~Dennen and Y.-t.~Huang,
\textit{``{Dual Conformal Properties of Six-Dimensional Maximal Super
  Yang-Mills Amplitudes}''},
\textsf{\doiref{10.1007/JHEP01(2011)140}{JHEP~1101,~140~(2011)}},
\texttt{\arxivref{1010.5874}{arxiv:1010.5874}}.

\bibitem{CaronHuot:2010rj}
S.~Caron-Huot and D.~O'Connell,
\textit{``{Spinor Helicity and Dual Conformal Symmetry in Ten Dimensions}''},
\texttt{\arxivref{1010.5487}{arxiv:1010.5487}}.

\bibitem{Cheung:2009dc}
C.~Cheung and D.~O'Connell,
\textit{``{Amplitudes and Spinor-Helicity in Six Dimensions}''},
\textsf{\doiref{10.1088/1126-6708/2009/07/075}{JHEP~0907,~075~(2009)}},
\texttt{\arxivref{0902.0981}{arxiv:0902.0981}}.

\end{thebibliography}
\bibliographystyle{nb}

\end{document}